\pdfoutput=1
\newcommand*{\ATLASLATEXPATH}{}
\documentclass[USenglish,texlive=2017,cernpreprint]{\ATLASLATEXPATH atlasdoc}
\usepackage[backend=bibtex]{\ATLASLATEXPATH atlaspackage} 
\usepackage{\ATLASLATEXPATH atlasbiblatex}

\usepackage[BSM]{\ATLASLATEXPATH atlasphysics}
\graphicspath{{logos/}{figures/}}

\usepackage{SUSY-2016-19-defs}

\newif\ifisconf
\isconffalse

\AtlasTitle{%
  Search for top squarks decaying to tau sleptons  \\
  in $p\mkern -1.5mu p$ collisions at \tttev  \\
  with the ATLAS detector
}

\author{The ATLAS Collaboration}

\AtlasRefCode{SUSY-2016-19}

\ifisconf
\AtlasNote{ATLAS-CONF-2017-079} %
\else
\AtlasNote{SUSY-2016-19} %
\fi

\PreprintIdNumber{CERN-EP-2018-024}

\AtlasDate{\today}

\AtlasJournal{PRD}
\AtlasJournalRef{\PRD 98, 032008 (2018)}
\AtlasDOI{10.1103/PhysRevD.98.032008}

\AtlasAbstract{%
  A search for direct pair production of top squarks in final states with two tau leptons, 
    $b$-jets, %
    and missing transverse momentum is presented.
  The analysis is based on proton--proton collision data at \tttev corresponding to an integrated luminosity of \intlumi 
    recorded with the ATLAS detector at the Large Hadron Collider in 2015 and 2016.
  Two exclusive channels 
    with either two hadronically decaying tau leptons or one hadronically and one leptonically decaying tau lepton
    are considered.
  No significant deviation from the Standard Model predictions is observed in the data.
  The analysis results are interpreted in terms of model-independent limits 
    and used to derive exclusion limits on the masses of the top squark $\tone$ and the tau slepton $\tauone$ 
    in a simplified model of supersymmetry with a nearly massless gravitino.
  In this model, masses up to $\massof{\tone}=\limittone$ and $\massof{\tauone}=\limittauone$ are excluded at \percent{95} confidence level.
}

\ifdef{\AtlasCoverSupportingNote}{ %

\AtlasCoverSupportingNote{Summer 2017 INT note}{https://cds.cern.ch/record/2231791}
\AtlasCoverCommentsDeadline{22nd February 2018}

\AtlasCoverEgroupEditors{atlas-SUSY-2016-19-editors@cern.ch}

\AtlasCoverEgroupEdBoard{atlas-SUSY-2016-19-editorial-board@cern.ch}

}{} %

\hypersetup{pdftitle={SUSY-2016-19},pdfauthor={The ATLAS Collaboration}}

\begin{document}

\maketitle

\ifisconf
\else
\tableofcontents 
\fi

\section{Introduction}
\label{sec:introduction}

Supersymmetry (SUSY)~\cite{Golfand:1971iw,Volkov:1973ix,Wess:1974tw,Wess:1974jb,Ferrara:1974pu,Salam:1974ig} (see Ref.~\cite{Martin:1997ns} for a review)
  extends the Standard Model (SM) with an additional symmetry
  that connects bosons and fermions, thereby providing answers to several of the open questions in the SM.
It predicts the existence of new particles 
  that have the same mass and quantum numbers as their SM partners
  but differ in spin by one half-unit.
Since no such particles have yet been observed,
  SUSY, if realized in nature, must be a broken symmetry, 
  allowing the supersymmetric partner particles to have higher masses than their SM counterparts.
In the model considered in this work, %
  the conservation of $R$-parity is assumed~\cite{Farrar:1978xj},
  so that the supersymmetric particles (sparticles) are produced in pairs,
  and the lightest supersymmetric particle (LSP) %
  is stable, 
  providing a viable candidate for dark matter.

This \article describes a search for SUSY in a benchmark scenario 
  motivated by gauge-mediated SUSY breaking~\cite{Dine:1981gu,AlvarezGaume:1981wy,Nappi:1982hm} 
  and natural gauge mediation~\cite{Asano2010}.
Assuming a mass spectrum for the sparticles
  that naturally avoids large fine-tuning~\cite{Barbieri:1987fn,deCarlos:1993yy}, %
  the scalar partner of the top quark (top squark) %
  is expected to be light. %
Furthermore, the scalar partner of the tau lepton (tau slepton) %
  is often the lightest charged slepton,
  motivating a search that focuses on final states with tau leptons.
In the benchmark scenario considered here, only three sparticles are assumed to be sufficiently light to be relevant for phenomenology at the Large Hadron Collider (LHC): 
  the lightest %
  top squark (\tone),
  the lightest tau slepton (\stauone), 
  and a nearly massless gravitino ($\gravitino$).

\begin{figure}[tbp]
  \centering
  \includegraphics[width=4.8cm]{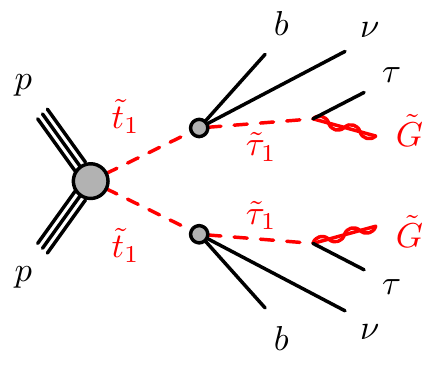} %
  \caption{\discussID{3}
    The simplified model for production and decay of supersymmetric particles considered in this analysis.
    The branching ratios are assumed to be \percent{100} in the decay mode shown, 
      both for the decay of the top squark as well as for the decay of the tau slepton.
    All sparticles not appearing in this diagram are assumed to be too massive to be relevant for LHC phenomenology.
    The top-squark decay vertex is drawn as a blob to indicate that the three-body decay is assumed to happen through an off-shell chargino.
  }
  \label{fig:stopstau_decay}
\end{figure}

The search strategy is optimized using a simplified model \cite{
    Alwall:2008ag,
    Alves:2011wf,
  } with this limited sparticle content.
The relevant parameters are the sfermion masses $\massof{\tone}$ and $\massof{\stauone}$.
The process is illustrated in \fig{fig:stopstau_decay}.
The top squark is directly pair-produced through the strong interaction.
Each top squark decays to a $b$-quark, a tau neutrino, and a tau slepton
  which in turn decays to a tau lepton and a gravitino.
The branching ratios for these decays are set to \percent{100}, %
  and the decays are assumed to be prompt.
The tau-slepton mixing matrix is chosen such that the tau slepton is an equal mix of the superpartners of the left- and the right-handed tau lepton.
Alternative scenarios with a neutralino \None as the LSP,
  which would suggest a high branching ratio of direct decays $\tone\to t\ninoone$,
  have been studied elsewhere~%
  \cite{
    SUSY-2014-07,     %
    Aaboud:2016lwz,   %
    Aaboud:2017nfd,   %
    Sirunyan:2017wif, %
    Sirunyan:2017xse, %
  }.

The search uses proton--proton ($pp$) collision data 
  collected with the ATLAS detector at a center-of-mass energy of \tttev in 2015 and 2016, 
  with a combined integrated luminosity of \intlumi.
A previous analysis considering the same three-body decay mode of the top squark to the tau slepton
  based on \ifb{20} of ATLAS data at \eighttev 
  set lower limits on the mass of the top squark \tone 
  of up to \oldlimittone~\cite{SUSY-2014-04}.
The combined LEP lower limit on the mass of the tau slepton,
  derived from searches for $\stau\to\tau\None$ decays and assuming the unification of gaugino masses,
  ranges between \GeV{87} and \GeV{96}
  depending on the assumed mass of the lightest neutralino~\cite{web:LEP}.
Models with small mass differences between the tau slepton and the lightest neutralino of up to approximately \GeV{10}
  are not excluded by the LEP experiments.
For a branching ratio $\stau \to \tau \None$ of \percent{100} and a massless \None,
  the lower limit on the tau-slepton mass is around \GeV{90}.
The limits published by the LHC experiments~\cite{SUSY-2013-14,SUSY-2014-05}
  obtained from models with direct production of tau sleptons 
  are not more stringent than those provided by LEP.

 Final states with two tau leptons can be classified into one of three channels,
depending on the decay modes of the tau leptons.
If both tau leptons decay hadronically, events
  belong to the ``had-had'' channel.
The ``lep-had'' channel refers to events
  in which one of the tau leptons decays leptonically and the other hadronically.
Final states where both tau leptons decay leptonically
  have the smallest branching ratio
  and are not considered,
  as studies showed that they would not contribute significantly to the sensitivity of the analysis. %

This \article is structured as follows.  %
\sec{sec:detector} gives a brief description of the ATLAS detector.
\Sec{sec:dataset} describes the recorded and simulated events used in the analysis,
  while \sec{sec:eventreconstruction} summarizes the reconstruction of physics objects such as leptons and jets
  and the kinematic variables used in the event selection.  
In \sec{sec:eventselection}, the selection used to obtain a signal-enriched event sample is described.
The background determination is described in \sec{sec:background_estimation},
  followed by a discussion of the methods used to derive the corresponding systematic uncertainties in \sec{sec:systematics}.  
\Sec{sec:results} presents the analysis results and their interpretation.
The \article concludes with a brief summary in \sec{sec:conclusion}.
\section{ATLAS detector}
\label{sec:detector}

The ATLAS detector~\cite{PERF-2007-01} is a multipurpose particle detector 
  with a forward--backward symmetric cylindrical geometry and nearly $4\pi$ coverage in solid angle.%
\footnote{
  ATLAS uses a right-handed coordinate system with its origin at the nominal interaction point (IP) in the center of the detector and the $z$-axis along the beam pipe. 
  The $x$-axis points from the IP to the center of the LHC ring, and the $y$-axis points upward. 
  Cylindrical coordinates $(r,\phi)$ are used in the transverse plane, 
    $\phi$ being the azimuthal angle around the $z$-axis. 
  The pseudorapidity is defined in terms of the polar angle $\theta$ as $\eta=-\ln\tan(\theta/2)$.
  When the mass of a particle cannot be neglected, 
    the rapidity $y=0.5 \ln\left[(E + p_z )/(E - p_z )\right]$ is used instead of the pseudorapidity $\eta$ to specify its direction.
}
It consists of, starting from the interaction point and moving outwards, an inner tracking detector, electromagnetic and hadronic calorimeters, and a muon spectrometer.
The inner tracking detector %
  covers the pseudorapidity range $\abseta < 2.5$
  and consists of silicon pixel, silicon microstrip, and transition radiation %
    detectors,
  immersed in a \unit[2]{T} axial magnetic field provided by a thin superconducting solenoid.
The insertable B-layer, the innermost layer of the silicon pixel detector,
  was added before the \tttev data-taking
  and provides high-resolution hits to improve the tracking and $b$-tagging performance~\cite{Capeans:1291633,CERN-LHCC-2012-009}.
The calorimeter system covers pseudorapidities $\abseta < 4.9$.
Electromagnetic energy measurements with high granularity are provided by lead/liquid-argon sampling calorimeters in the region $\abseta < 3.2$.
Hadronic calorimetry is provided by sampling calorimeters 
  with scintillator tiles and steel absorbers within $\abseta < 1.7$ 
  and with copper/liquid-argon for $1.5 < \abseta < 3.2$.
The forward regions are instrumented with sampling calorimeters using liquid-argon as the active medium for both the electromagnetic and hadronic calorimetry.
The muon spectrometer features three large superconducting toroid magnets with eight coils each,
  precision-tracking detectors in the region $\abseta < 2.7$,
  and fast, dedicated chambers for triggering in the region $\abseta < 2.4$.
Collision events are selected for recording by a two-stage trigger system,
  which has been upgraded for the run at \tttev~\cite{TRIG-2016-01}.
It consists of a hardware-based trigger as the first level, %
  followed by the software-based high-level trigger, %
  which is able to run reconstruction and calibration algorithms similar to those used offline,
  reducing the event rate to about \unit[1]{kHz}.

\section{Dataset and simulation}
\label{sec:dataset}

The dataset analyzed in this \article was recorded with the ATLAS detector
  from $pp$ collisions delivered by the Large Hadron Collider at CERN in %
  2015 and 2016
  at a center-of-mass energy of \tttev~\cite{PhysRevAccelBeams.20.081003}. %
Collision events are selected with triggers on electrons or muons in the lep-had channel, and on missing transverse momentum or two hadronic tau leptons in the had-had channel.
The total integrated luminosity of the dataset 
  after the application of data-quality requirements 
  that ensure that all subdetectors are functioning normally
  is \ifb{36.1} with an uncertainty of \percent{3.2}.
The uncertainty was derived, following a methodology similar to that detailed in Ref.~\cite{DAPR-2013-01}, 
  from a preliminary calibration of the luminosity scale using $x$--$y$ beam-separation scans performed in August 2015 and May 2016.

\begin{table}[tbp]
  \centering                                                        
  \newcommand{\maybeugly}{@{\hspace{2\tabcolsep}}}
  \newcommand{\zr}[1]{\multirow{2}{*}{#1}}
  \begin{tabular}{lll @{\hspace{2mm}} ll @{\hspace{2mm}} l}
    \toprule                                                        
    Process     & Matrix element        & PDF set          & Parton shower       & PDF set          & Tune          \\
    \midrule %
    \ttbar      & \POWHEGBOX    v2      & CT10             & \PYTHIA     6.428   & CTEQ6L1          & \Perugia 2012 \\ 
    Single-top  & \POWHEGBOX    v1      & CT10             & \PYTHIA     6.428   & CTEQ6L1          & \Perugia 2012 \\ %
    $\ttbar H$  & \MGaMC        2.2.2   & CT10             & \HERWIGpp   2.7.1   & CTEQ6L1          & UE-EE-5       \\
    $\ttbar V$  & \MGaMC        2.3.3   & NNPDF3.0 NLO     & \PYTHIA     8.210   & NNPDF2.3 LO      & A14           \\
    $tWZ$       & \MGaMC        2.3.2   & NNPDF3.0 NLO     & \PYTHIA     8.210   & NNPDF2.3 LO      & A14           \\
    $tZ$        & \MGaMC        2.2.1   & CTEQ6L1          & \PYTHIA     6.428   & CTEQ6L1          & \Perugia 2012 \\ %
    Multi-top   & \MGaMC        2.2.2   & NNPDF2.3 LO      & \PYTHIA     8.186   & NNPDF2.3 LO      & A14           \\
    \midrule %
    \Vjets      & \SHERPA       2.2.1   & NNPDF3.0 NNLO    &                     &                  &               \\
    $VV$ (1)    & \SHERPA       2.2.1   & NNPDF3.0 NNLO    &                     &                  &               \\
    $VV$ (2)    & \SHERPA       2.1.1   & CT10             &                     &                  &               \\
    $VVV$       & \SHERPA       2.2.2   & NNPDF3.0 NNLO    &                     &                  &               \\
    \midrule %
    \zr{SUSY}   & \MGaMC                & \zr{NNPDF2.3 LO} & \pythia 8.186,      & \zr{NNPDF2.3 LO} & \zr{A14}      \\
                &  2.2.3 and  2.3.3     &                  & 8.205 or 8.210      &                  &               \\
    \bottomrule
  \end{tabular}
  \caption{
    Overview of the simulation codes, parton distribution function (PDF) sets and parameters 
      used to simulate the Standard Model background processes
      and the supersymmetric signal process (SUSY).
    \MGMCatNLO is abbreviated as \MGaMC.
    Corresponding references are given in the text.
  }
  \label{tab:MC_gen}
\end{table}
 
Monte Carlo (MC) simulation was used to generate samples of collision events, %
  which model the expected kinematics of the supersymmetric signal 
  and allow the prediction of the contributions from the various SM background processes.
The MC generators, parton distribution function (PDF) sets and parameters %
  used to simulate the Standard Model background processes
  and the supersymmetric signal process of the simplified model
  are summarized in \tab{tab:MC_gen}.
Additional MC samples are used to estimate systematic uncertainties, as described in \sec{sec:systematics}.
 Data-driven methods are used to augment the accuracy of the simulation-based estimates for the major background processes
  (cf.\ \sec{sec:background_estimation}).

Signal samples were generated from leading-order (LO) matrix elements (ME) %
  with \textsc{MadGraph5}-\aMCatNLO %
  v2.2.3 and v2.3.3~\cite{Alwall:2014hca}
  interfaced to \pythia 8.186, 8.205 or 8.210~\cite{Sjostrand:2007gs,Sjostrand:2014zea} %
  with the ATLAS 2014 (A14)~\cite{ATL-PHYS-PUB-2014-021} %
  set of tuned parameters (tune) %
  for the modeling of the parton showering (PS), %
  hadronization and underlying event. 
The matrix element calculation was performed at tree level and includes the emission of up to two additional partons. 
The PDF set
  used for the generation was NNPDF2.3 LO~\cite{Ball:2012cx}. %
The ME--PS matching was done using the CKKW-L~\cite{Lonnblad:2012ix} prescription, 
  with the matching scale set to one quarter of the top-squark mass
  in accordance with the recommendations. %
Signal cross sections were calculated to next-to-leading order (NLO) in the strong coupling constant, 
  adding the resummation of soft gluon emission at next-to-leading logarithmic accuracy%
  ~\cite{Beenakker:1997ut,Beenakker:2010nq,Beenakker:2011fu}. 
Production of top-quark pairs and of single top quarks in the $s$- and $t$-channel or associated with $W$~bosons 
  was simulated at NLO with \POWHEGBOX~%
  \cite{
    Alioli2010,      %
    Re:2010bp,       %
    Frixione:2007nw, %
    Frederix:2012dh, %
    Alioli:2009je,   %
  } 
  interfaced to \PYTHIA 6.428~\cite{Sjostrand:2006za}
  for the parton shower, hadronization, and underlying event, %
  using the CT10 PDF set~\cite{Lai:2010vv} %
  in the matrix element calculations
  and the CTEQ6L1 PDF set~\cite{Pumplin:2002vw} %
  with the \Perugia 2012 tune~\cite{Skands:2010ak} for the parton shower and underlying event.
Associated production of top-quark pairs and Higgs bosons was simulated at NLO with \MGMCatNLO~\cite{Alwall:2014hca} interfaced to \HERWIGpp~2~\cite{Bahr:2008pv,Bellm:2013hwb},
  using the UE-EE-5 tune~\cite{Gieseke:2012ft}. %
For \ttV, where $V$ is a $W$ or $Z$ boson, and $tWZ$ production at NLO,
  \MGMCatNLO with the NNPDF3.0 NLO PDF set~\cite{Ball:2014uwa} %
  and \PYTHIA 8.210~\cite{Sjostrand:2014zea} %
  were used. 
Finally, production of $tZ$ and three or four top quarks (multi-top) was simulated at LO with \MGMCatNLO and \PYTHIA.
The EvtGen program~\cite{LANGE2001152} was used for all samples with top quarks and the signal samples to model the properties of the bottom- and charm-hadron decays.

Drell--Yan production of charged and neutral leptons, $Z/\gamma^* \to \ell^+\ell^-$ and $Z \to \nu\bar\nu$,
  and leptonic decays of $W$~bosons, $W\to\ell\nu$,
  in association with jets (\Vjets) were simulated~\cite{ATL-PHYS-PUB-2017-006} with \SHERPA~\cite{Gleisberg:2008ta},
  using the \SHERPA parton shower~\cite{Schumann:2007mg} and a dedicated tuning developed by the \SHERPA authors. %
\SHERPA was also used for the simulation of diboson production ($VV$)
  and leptonic decays of triboson production ($VVV$)~\cite{ATL-PHYS-PUB-2017-005}.
The diboson samples include one set of tree-induced processes with dileptonic and semileptonic decays, $VV$~(1), %
  and a second set with electroweak $VVjj$ production %
  and loop-induced production with leptonic decays, $VV$~(2). %

All simulated background events were passed through a full \GEANTiv~\cite{AGOSTINELLI2003250} simulation of the ATLAS detector~\cite{SOFT-2010-01}.  
For signal events, a fast detector simulation was used,
  which is based on a parameterization of the performance of the electromagnetic and hadronic calorimeters~\cite{ATL-PHYS-PUB-2010-013} 
  and on \GEANTiv for all other detector components.
The same algorithms were used for the reconstruction of physics objects 
  in simulated signal and background events
  and in collision data. %
Agreement between simulated events and collision data is improved by weighting the simulated events 
  to account for differences in the 
    lepton trigger, 
    reconstruction, 
      identification and 
      isolation efficiencies,
    $b$-tagging efficiency, and
    jet-vertex-tagging efficiency,
  using correction factors derived in dedicated studies.

The effect of additional $pp$ interactions in the same and nearby bunch crossings (``pileup'') %
  was taken into account by overlaying the hard-scattering process
  with soft $pp$ interactions generated with \PYTHIA 8.186 using the A2 tune~\cite{ATL-PHYS-PUB-2012-003}
  and the MSTW2008LO PDF set~\cite{Martin:2009iq}.
Simulated events were reweighted to make the distribution of the average %
  number of simultaneous $pp$ collisions match that of the recorded dataset.

\section{Event reconstruction}
\label{sec:eventreconstruction}

The data recorded in collision events are processed 
  to reconstruct and identify physics objects needed for the event selection, and to reject events of insufficient quality.
Candidate events are required to have a reconstructed vertex~\cite{ATL-PHYS-PUB-2015-026}
  with at least two associated tracks with a transverse momentum $\pT > \MeV{400}$. %
If there are several such vertices, the one with the largest scalar sum of $\pT^2$ of its associated tracks is used as the primary collision vertex.

Jets are reconstructed from topological energy clusters in the calorimeters~\cite{ATL-PHYS-PUB-2015-036,PERF-2014-07} %
  using the \antikt algorithm~\cite{Cacciari:2008gb} %
  with radius parameter $R=0.4$ and are calibrated to the hadronic scale, %
  accounting for the impact of pileup in the event. %
The calibration is improved with the global sequential correction scheme~\cite{Aaboud:2017jcu}. %
Jets with $\pT>\GeV{20}$ and $\abseta<2.8$ are retained. %
In addition, jets need to fulfill %
  basic quality criteria; an event is discarded if any selected jet does not meet these criteria~\cite{ATLAS-CONF-2015-029}.
Pileup is suppressed further by rejecting jets with $\pT<\GeV{60}$ and $\abseta<2.4$
  if the output of a jet-vertex-tagging algorithm~\cite{PERF-2014-03} shows their origin is not compatible 
  with the primary vertex.

A multivariate discriminant based on track impact parameters and reconstructed secondary vertices~\cite{PERF-2012-04, ATL-PHYS-PUB-2016-012} is employed 
  to identify jets with $\abseta < 2.5$ resulting from the hadronisation of $b$-quarks~(\bjets). %
The chosen working point has a $b$-tagging efficiency of \percent{77}
  and rejection factors of 134, 6, and 22, for light-quark and gluon jets, $c$-quark jets, and hadronically decaying tau leptons, $\tauh$, respectively,
  as evaluated on a simulated sample of \ttbar events.

A dedicated algorithm is used to reconstruct $\tauh$ candidates and match them to a primary vertex.
This is seeded from jets reconstructed with the \antikt algorithm 
  with a radius parameter $R = 0.4$ and fulfilling $\pt > \GeV{10}$ and $\abseta < 2.5$~\cite{ATL-PHYS-PUB-2015-045}.
Only the visible part of the \tauh decay is reconstructed.
An energy calibration derived independently of the jet energy scale is applied to the reconstructed $\tauh$~\cite{PERF-2013-06}.
The analysis uses $\tauh$ candidates with $\pt>\GeV{20}$ and $\abseta < 2.5$, 
  excluding the calorimeter transition region $1.37 < \abseta < 1.52$ because of its larger uncertainty in jet direction measurements.
 The $\tauh$ candidates are required to have one or three associated tracks (prongs) and a total track charge of $\pm1$.  
A discriminant obtained from a boosted decision tree is used to reject jets that do not originate from a 
hadronically decaying tau lepton, 
  with a working point yielding a combined tau reconstruction and identification efficiency of \percent{55} (\percent{40}) for 1-prong (3-prong) $\tauh$~\cite{ATLAS-CONF-2017-029}.
A looser set of identification criteria, called ``AntiID'', are used
for the background estimate using the fake-factor method, as described in \sec{sec:fake-factor-method}.

Two sets of identification criteria are defined for electrons and muons: 
  the \emph{baseline} criteria are used for lepton vetoes and the overlap removal procedure described below, 
  while \emph{signal} criteria are used when the event selection requires the presence of a lepton.

Electron candidates are reconstructed from energy clusters in the electromagnetic calorimeter matched to tracks in the inner tracking detector.
Baseline electrons must satisfy a loose likelihood-based identification~\cite{ATLAS-CONF-2016-024, ElectronPerformance2017}
  and have $|\eta_\text{cluster}|<2.47$ and $\pT>\GeV{10}$.
Signal electrons must have $\pt > \GeV{25}$ and satisfy the tight likelihood-based quality criteria.
Isolation requirements using calorimeter- and track-based information are applied %
  that provide \percent{95} efficiency for electrons with $\pt = \GeV{25}$, rising to \percent{99} efficiency at $\pt = \GeV{60}$ in $Z\to ee$ events.
In addition, signal electrons must fulfill requirements on the transverse impact parameter significance ($\left|d_0\right|/\sigma(d_0) < 5$)
  and the longitudinal impact parameter ($\left|z_0 \sin(\theta)\right| < \unit[0.5]{mm}$).

The muon reconstruction combines tracks recorded in the muon system with those reconstructed in the inner tracking detector.
Baseline muons must have $\pt > \GeV{10}$ and $\abseta < 2.7$ and fulfill medium quality criteria~\cite{MuonPerformance2016}. %
Signal muons must satisfy $\pt > \GeV{25}$ and $\abseta < 2.5$
  and isolation requirements similar to those for signal electrons
  as well as requirements on the track impact parameters ($\left|d_0\right|/\sigma(d_0) < 3$ and $\left|z_0 \sin(\theta)\right| < 0.5$ mm).

The jet and lepton reconstruction algorithms described above work independently of each other and may therefore assign the same detector signature to multiple objects.  
A sequence of geometrical prescriptions is applied to resolve ambiguities by removing objects.  
In particular, $\tauh$ candidates near electrons or muons 
  ($\Delta R_y\equiv\sqrt{(\Delta\phi)^2+(\Delta y)^2} < 0.2$%
  ) %
  are discarded as part of this procedure.
No jet is allowed near an electron or a muon: %
  for $\Delta R_y<0.2$, the jet is removed, while for $0.2<\Delta R_y<0.4$, the lepton is removed instead.
The missing transverse momentum \vecmet is defined as the negative vector sum of the transverse momenta of all identified physics objects 
  (electrons, photons, muons, tau leptons, jets) %
  and an additional soft-track term. 
The soft-track term is constructed from all tracks that are not associated with any reconstructed physics object
  but are associated with the identified primary collision vertex~\cite{ATL-PHYS-PUB-2015-023,ATL-PHYS-PUB-2015-027}.
In this way, the missing transverse momentum benefits from the calibration of the identified physics objects,
  and remaining energy deposits are included in a pileup-insensitive manner.
Frequently, only the magnitude $\met \equiv |\vecmet|$ is used.

\subsection{Analysis variables}
\label{sec:analysisvariables}

Besides basic kinematic quantities, the variables described below are used in the event selection.

The transverse mass \mT %
  is computed from the transverse momentum of a lepton $\ell$ and the missing transverse 
  momentum %
  in the event:
\begin{equation*}
  \mT=\sqrt{2\, \met \,\pT,_\ell\cdot\left( 1-\cos \left(\Delta \phi \left(\vecmet,\vec\pT,_\ell\right) \right)\right) }, %
\end{equation*}
where $\vec\pT,_\ell$ is the lepton's transverse momentum.  In \Wjets events, the \mT distribution has a cutoff near the $W$-boson mass $\massof{W}$.

The stransverse mass \mTtwo~\cite{
  arxiv:LesterMT2,       %
  Barr:2003rg,           %
  Lester:2014yga,        %
}
is employed in this analysis to reject the top-pair background.
It is a generalization of the transverse mass for final states with two invisible particles.
It assumes two identical particles that decay to one visible and one invisible product each, 
  and provides an upper bound on the mother particle's mass.
This is achieved by considering all possible ways to distribute the measured \vecmet between the invisible particles of the assumed decay.

Here, \mTtwo is constructed using the leptons as the visible particles.  
The \vecmet is assumed to stem from a pair of neutrinos,
  \ie the mass hypothesis for the invisible particles is set to zero in the computation of \mttwo.  %
The resulting variable is a powerful discriminant against  background events from \ttbar or $WW$ production, 
  as it is bounded from above by $\massof{W}$ for these, while signal events do not respect this bound.

Furthermore, the invariant mass $m(\ell_1,\ell_2)$ of the two reconstructed leptons (including $\tauh$), as well as
\HT, defined as the scalar sum of the \pt of the two leading jets, is used.

\section{Event selection}
\label{sec:eventselection}

The event selection starts from preselections that are similar for the lep-had and had-had channels, 
 differing only in the choice of event triggers and the required numbers of reconstructed tau leptons and light leptons,
  \ie electrons and muons. %
Prompt light leptons are not distinguished from light leptons originating from decays of tau leptons.
Therefore, in the background estimates, 
  processes with prompt light leptons contribute in the same way as processes with leptonic decays of tau leptons.
The event selections for the two channels are mutually exclusive.
The channels can therefore be statistically combined in the interpretation of the results.

\subsection{Preselection}
\label{sec:presel}

\begin{table}[ht]
  \centering
  \begin{tabular}{l l l }
  \toprule                                                     
    Preselection & lep-had                               & had-had                                             \\
  \midrule                                                                                                                
    Trigger & single-electron or single-muon trigger   & \met or di-tau trigger                                           \\[+1mm]
    \multirow{2}{*}{Leptons} & exactly one $\tauh$ + one signal electron or muon      & exactly two $\tauh$                                                \\
    & no additional baseline electron or muon or \tauh   & no baseline electron or muon                                     \\
    Trigger-related & \multirow{2}{*}{$\pt(e,\mu) > \GeV{27}$}  & $\met > \GeV{180}$ or \\
    \multicolumn{1}{r}{requirements} &  & $\pt(\tau_{1,2}, \text{jet}_1) > \GeV{50,40,80}$   \\
  \midrule %
    $\pt(\text{jet}_2)$ &  $> \GeV{26}$                 & $ > \GeV{20}$                                   \\
    $\pt(\tau_1)$ &  $> \GeV{70}$                        & $> \GeV{70}$                                         \\
    $n_{\bjet}$ &  $ \geq 1$                           & $ \geq 1$                                               \\
  \bottomrule
  \end{tabular}
  \caption{\discussID{14}
    Preselections in the lep-had and had-had channel.
    The leading (subleading) objects are referred to using indices, \eg $\text{jet}_1$ ($\text{jet}_2$),
      and $\tau_1$ ($\tau_2$) refers to the leading (subleading) \tauh.
  }
  \label{tab:presel_compare}
\end{table}

The preselection requirements for the two channels are summarized in \tab{tab:presel_compare}.
In the lep-had channel, events selected by single-electron or single-muon triggers are used.  
The had-had channel uses a logical OR of an \met trigger 
  and a combined trigger selecting events with two tau leptons and one additional jet at the first trigger level. %
The preselection adds suitable requirements to avoid working in the turn-on regime of the trigger efficiency.
For events selected by the single-lepton triggers, 
  the \pt of the light lepton is required to be at least \GeV{27}.
For events selected by the \met trigger, \met needs to exceed \GeV{180}, 
  and for events selected by the combined trigger,
  the requirements are at least \GeV{50} (\GeV{40}) for the \pt %
  of the leading (subleading) \tauh, 
  and $\pt > \GeV{80}$ for the leading jet, %
  where leading refers to the object with the largest transverse momentum. %
The trigger efficiencies, 
  which are used to compute scale factors that correct for small differences between simulation and collision data,
  are measured as a function of the properties of leptons reconstructed offline, so these leptons are matched to 
  the leptons reconstructed in the trigger.

All candidate events must have at least two jets with \pt larger than \GeV{26} (\GeV{20}) in the lep-had (had-had) channel. %
For the lep-had channel, the preselection requires exactly one $\tauh$, exactly one signal electron or muon, and no further baseline leptons.  
For the had-had channel, exactly two $\tauh$ are required, and no baseline light leptons must be present.
No requirement on the electric charge of the leptons is applied in the preselection,
  as both events with opposite-charge and events with same-charge lepton pairs are used in the analysis.
In all regions of both the lep-had and had-had channels,
  the leading hadronically decaying tau lepton
  must have $\pt > \GeV{70}$.
In addition, %
  events are required to have at least one $b$-tagged jet ($n_{\bjet} \geq 1$).

\subsection{Signal selections}
\label{sec:signal_selections}

Two signal regions (SRs) %
  are defined, one for the lep-had channel and one for the had-had channel.
Both SR selections are based on the preselection described above,
  where in addition the lepton pair has to have opposite electric charge,
  as same-charge lepton pairs are not predicted by the signal model.
They were optimized to give the largest sensitivity to the targeted signal model %
  in terms of the discovery $p$-value computed using a ratio of Poisson means~\cite{2008NIMPA.595..480C,COUSINS2010388}. %

The variables with the best discrimination power between signal and background 
  are the missing transverse momentum and stransverse mass.
The optimal selection thresholds for these two variables are different in the two channels.
In the lep-had (had-had) channel,
  the signal selection requires $\mttwo > \GeV{100}$ (\GeV{80})
  and $\met > \GeV{230}$ (\GeV{200});
  the lep-had selection needs slightly higher thresholds to achieve the same discrimination power between signal and background.
A summary of the SR definitions is included in the last column 
  of \tabs{tab:def_CRVR_lephad} and~\ref{tab:def_CRVR_hadhad}
  for the lep-had and had-had channels, respectively.

\section{Background estimation}
\label{sec:background_estimation}

The general strategy for estimating the SM background in this analysis 
  is to develop dedicated control regions (CRs) %
  for the most important background contributions.
These CRs provide %
  data-driven constraints on the overall normalization of the respective background processes,
  whereas the shape of the kinematic distributions is taken from simulation. %
A maximum-likelihood fit is performed for all control-region yields simultaneously
  in order to obtain the normalization factors. %
The normalization factors from this background fit %
  are then extrapolated using simulation to obtain the expected yields in the signal region.
Therefore, all control-region selections must be mutually exclusive,
  with respect to each other as well as to the signal regions.
The correctness of the extrapolation is checked in additional selections called validation regions (VRs), %
  which cover the intermediate range in \mttwo between the control and the signal regions, without overlapping either.

The targeted final state has two tau leptons, two $b$-quarks %
  and missing transverse momentum. %
The dominant SM background process with this signature is pair production of top quarks. %
This background process can contribute in two different ways.
In the first case, the objects from the top-quark decays are correctly reconstructed.
One of the $W$ bosons from the top-quark decays yields a hadronically decaying tau lepton;
  the other $W$ boson decays to a light lepton in the lep-had channel, either directly or through a tau-lepton decay,
  or to a second hadronically decaying tau lepton in the had-had channel.
In the second case,
  the background events contain a fake tau lepton,
  \ie an object which is not a tau lepton, most often a jet or an electron, but reconstructed as a hadronically decaying tau lepton. %
The probability of falsely identifying a jet or an electron as a tau lepton is only of the order of a few percent,
  but the branching ratio of $W$ bosons to jets or electrons is larger than that to hadronically decaying tau leptons.
Moreover, the requirement on \mttwo is more efficient in rejecting \ttbar~events with real tau leptons.
Therefore, \ttbar~events with fake tau leptons dominate after applying the signal-region selection requirements.
As the nature and quality of the modeling in simulation of these two background components from \ttbar~events may be very different,
  they are treated as separate background components in the following.
The CRs and methods used to estimate the background from \ttbar~events are introduced in \secs{sec:CRlephad} and~\ref{sec:CRhadhad}.
The contribution of events with a real tau lepton and a fake light lepton is expected to be negligible 
  due to the small misidentification probabilities for light leptons.

Subdominant contributions to the SM background come from diboson production,
  where often a jet is falsely identified as originating from a $b$-hadron decay,
  or \ttbar production in association with a vector boson,
  where most often the additional vector boson is a $Z$ boson that decays to neutrinos.
The CRs for these background processes
  are based on a selection of events with light leptons rather than hadronically decaying tau leptons,
  in order to obtain good purity and enough events in the CRs. %
Common normalization factors for the lep-had and had-had channels are derived.
These CRs are defined in \sec{sec:CRcommon}.

Finally, smaller contributions come from vector-boson production 
  (\Wjets and \Zjets, collectively denoted by \Vjets) and single-top production.
Multi-top, triboson production,
  and \ttbar production in association with a Higgs boson
  contribute very little to the signal regions 
  and are therefore summarized under the label ``others'' in the following. 
The contributions of all of these are estimated directly from simulation 
  and normalized to the generator cross section 
    for triboson production~\cite{ATL-PHYS-PUB-2016-002}
    and multi-top production, %
  and higher-order cross-section calculations for \Vjets, $\ttbar H$ and single-top production~\cite{
  ATL-PHYS-PUB-2016-003, %
  ATL-PHYS-PUB-2017-006, %
  ATL-PHYS-PUB-2016-005, %
  Aliev:2010zk, Kant:2014oha, Kidonakis:2010ux, Kidonakis:2013zqa, %
  }. 
Contributions from multi-jet events are not relevant for the analysis,
  as was verified using data-driven methods.
The multi-jet background is therefore neglected.

One signal benchmark point was chosen to illustrate the behavior of the signal in comparison to the background processes in kinematic distributions.
The mass parameters for this benchmark point are $\massof{\tone} = \GeV{1100}$ and $\massof{\stauone} = \GeV{590}$.
A larger mass-splitting between the top squark and the tau slepton
  yields more-energetic $b$-tagged jets in the final state, %
  whereas a higher tau-slepton mass yields tau leptons with higher transverse momentum. %
As both the top squark and the tau slepton have invisible particles among their decay products,
 the \met spectrum does not depend strongly on the mass of the intermediate particle, the tau slepton.

\subsection{Lep-had channel}
\label{sec:CRlephad}

\newcommand{\filler}[1]{#1} %

\begin{table}[ht]
  \centering
  \begin{tabular}{p{1.7cm}rrrrr}
    \toprule
    Variable               & \CRLHttReal       & \VRLHttReal           & \VRLHttFakeOS          & \VRLHttFakeSS \kern-1.2em  & SR LH         \\
    \midrule                                                                                  
    Charge($\ell$,$\tauh)$ &  opposite         & opposite              & opposite               & same            & opposite      \\
    \mttwoLT               &   $< \GeV{60}$    & \GeVrange{60}{100}    & \GeVrange{60}{100}     & $ >\GeV{60}$    & $>\GeV{100}$  \\
    \met                   &  $> \GeV{210}$    & $> \GeV{210}$         & $ >\GeV{150}$          & $ >\GeV{150}$   & $>\GeV{230}$  \\
    $\mT(\ell)$            &  $> \GeV{100}$    & $> \GeV{100}$         & $ <\GeV{100}$          & ---             & ---           \\
    \mLT                   &   ---             & ---                   & $>\GeV{60}$            & ---             & ---           \\      
    \bottomrule
  \end{tabular}
  \caption{\discussID{10}
    Definitions of the \ttbar control and validation regions and the signal region in the lep-had channel. 
    \DashNoCut
    \BracketRange
    A common preselection as given in \tab{tab:presel_compare} for the lep-had channel is applied.
  }
  \label{tab:def_CRVR_lephad}
\end{table}

The contribution of background events with real hadronically decaying tau leptons in the lep-had channel is estimated from simulation.
For top-quark pair production,
  the shape of the distribution of the observables is taken from simulation 
  but the overall normalization is derived from a dedicated CR.    
For events with fake tau leptons,
  it is difficult to design a CR with sufficiently high event yields and purity.
Moreover, the estimate of this background from simulation does not agree with the observed data in the VRs.
Therefore, the background estimate for events with fake tau leptons
  is derived using a data-driven method called the fake-factor method, which is discussed below.
  
The CR and three VRs enriched in top-quark events or events with fake tau leptons are defined in \tab{tab:def_CRVR_lephad}.
As explained above, %
  the CR and VRs cover a lower \mttwo range,
  with the VRs located between the CR and the SR 
  to check the extrapolation in this variable.
In all of these regions,
  the preselection %
  requirements for the lep-had channel from \tab{tab:presel_compare} are applied.
In the opposite-charge regions,
  the transverse mass $\mT(\ell)$ of the light lepton and the missing transverse momentum %
  is used 
  to separate \ttbar~events with real tau leptons from those with fake tau leptons.
Events with top-quark pairs,
  where one of the top quarks decays to a light lepton and the other decays hadronically,
  and a jet from the hadronic $W$-boson decay is misidentified as the tau lepton, 
  yield mostly small values of \mt.
In these events, there is only one neutrino (from the leptonic $W$-boson decay), %
  so the transverse mass has an endpoint near the $W$-boson mass.
Events where both the light lepton and the hadronically decaying tau lepton are real involve more neutrinos, 
  leading to tails of the \mt~distribution that go beyond this endpoint.
The extrapolation from the control region to the signal region is performed in \mttwo,
  which is correlated with \mt,
  but the validation regions cover the full \mt range
  so that any potential bias from the correlation of \mt and \mttwo would be visible there.
The requirement on \mLT is added to improve the purity of the VR.

The purity in the respective targeted background process %
  is about 
  \percent{74} in \CRLHttReal, 
  \percent{70} in \VRLHttReal, and
  \percent{43} in \VRLHttFakeOS.
As the purity of \VRLHttFakeOS in \ttbar~events with fake tau leptons is low, %
  an additional validation region, \VRLHttFakeSS, with a same-charge requirement is defined.
The same-charge requirement is very efficient in rejecting events where both leptons are real 
  and originate from the $W$~bosons in a \ttbar~event.
The correlation between the charge of a jet misidentified as a tau lepton and the charge of the light lepton in \ttbar~events is much smaller;
  thus, events with fake tau leptons are more likely to pass the same-charge selection, %
  yielding a purity of \percent{91} in \VRLHttFakeSS.
\begin{figure}
  \centering
  \includegraphics[width=0.5\textwidth]{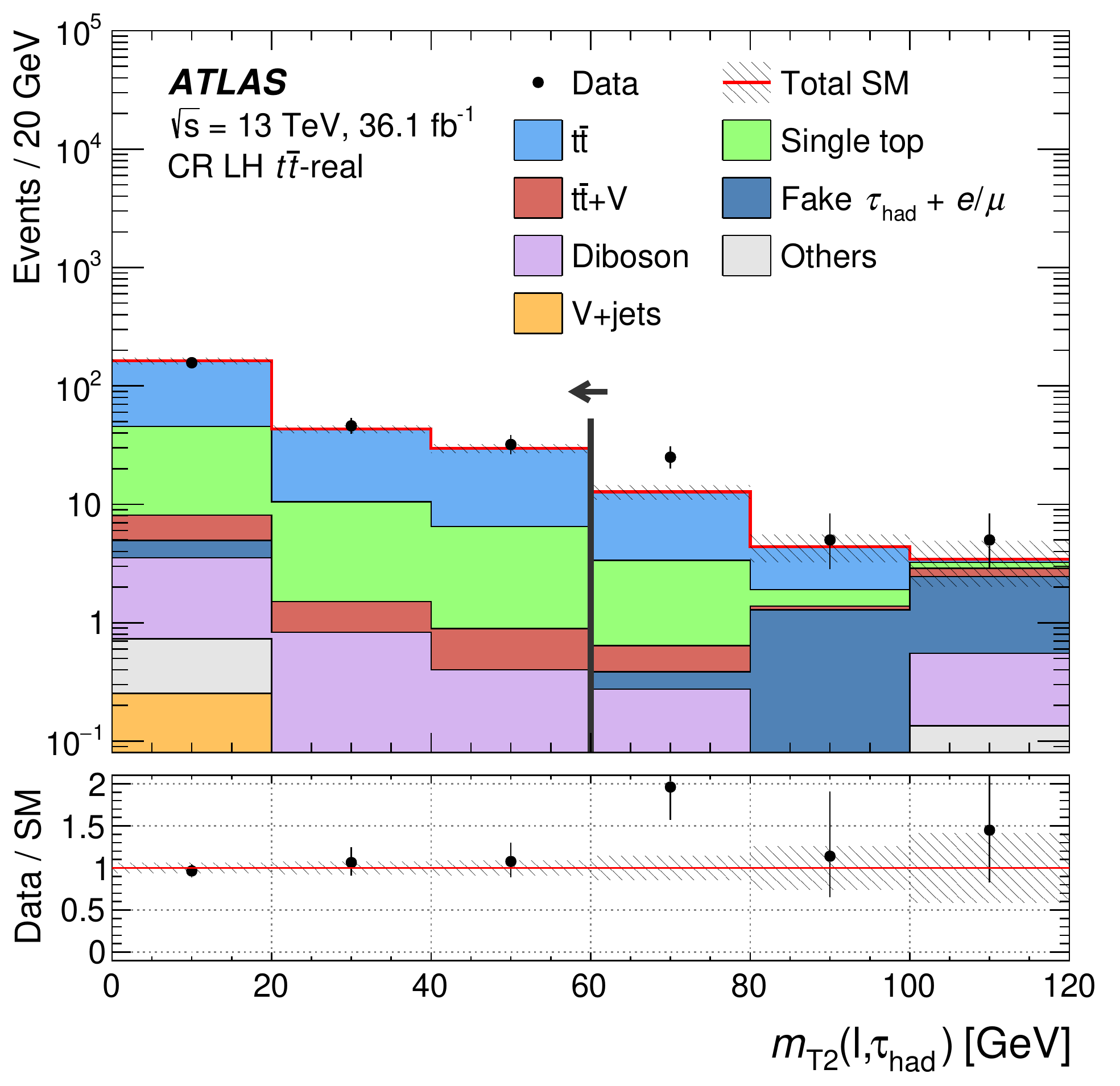}%
  \includegraphics[width=0.5\textwidth]{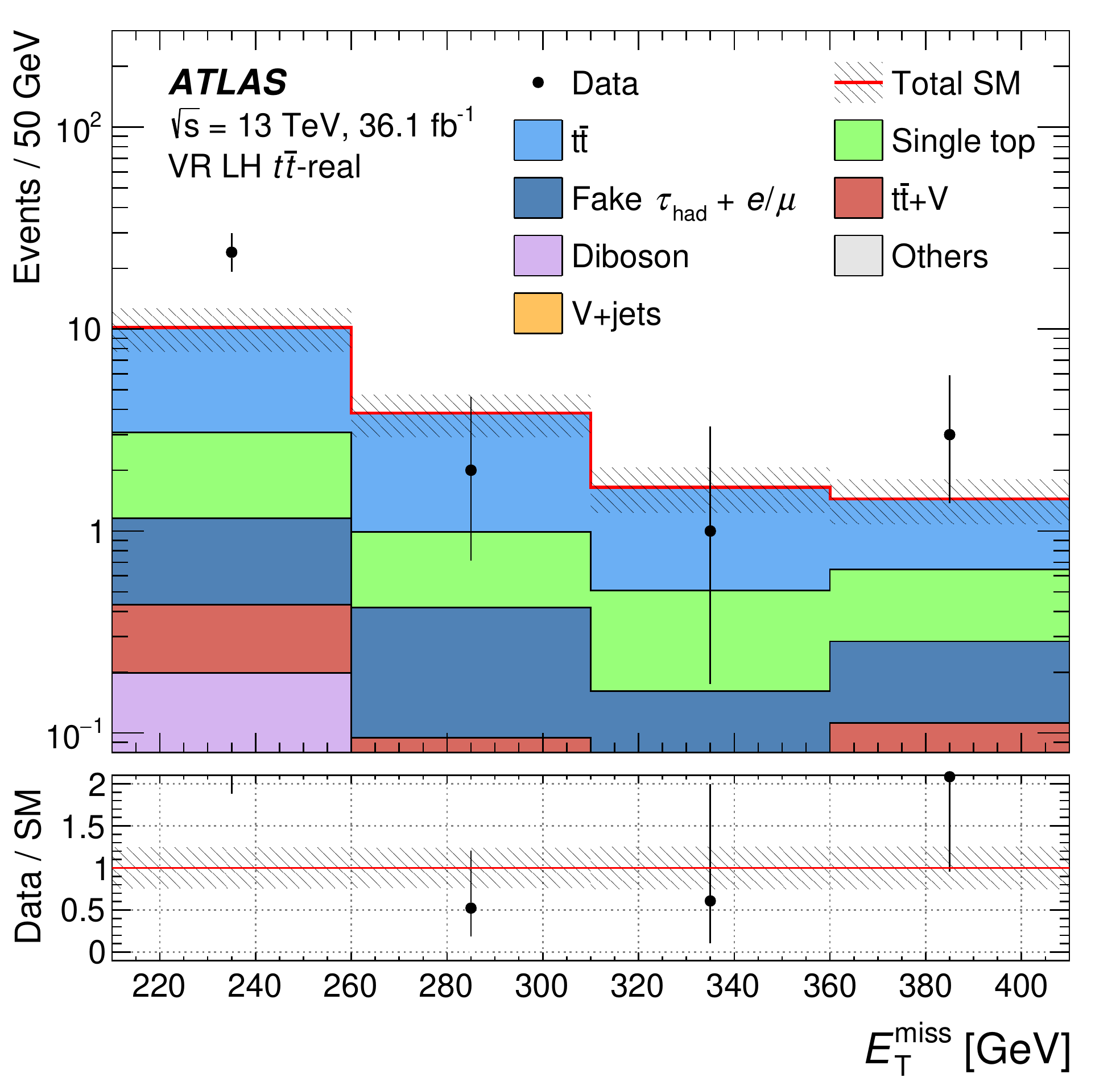}
  \includegraphics[width=0.5\textwidth]{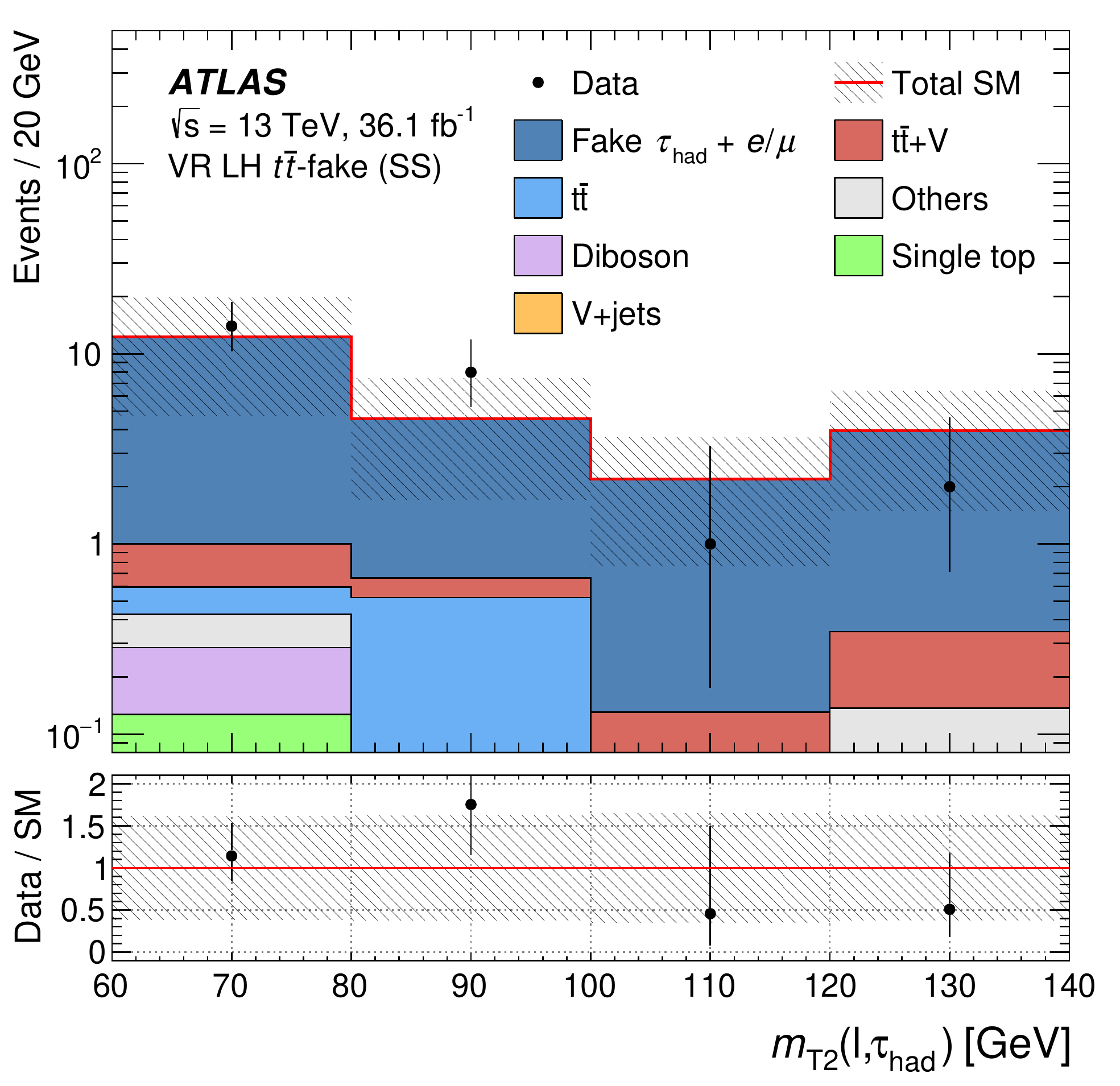}%
  \includegraphics[width=0.5\textwidth]{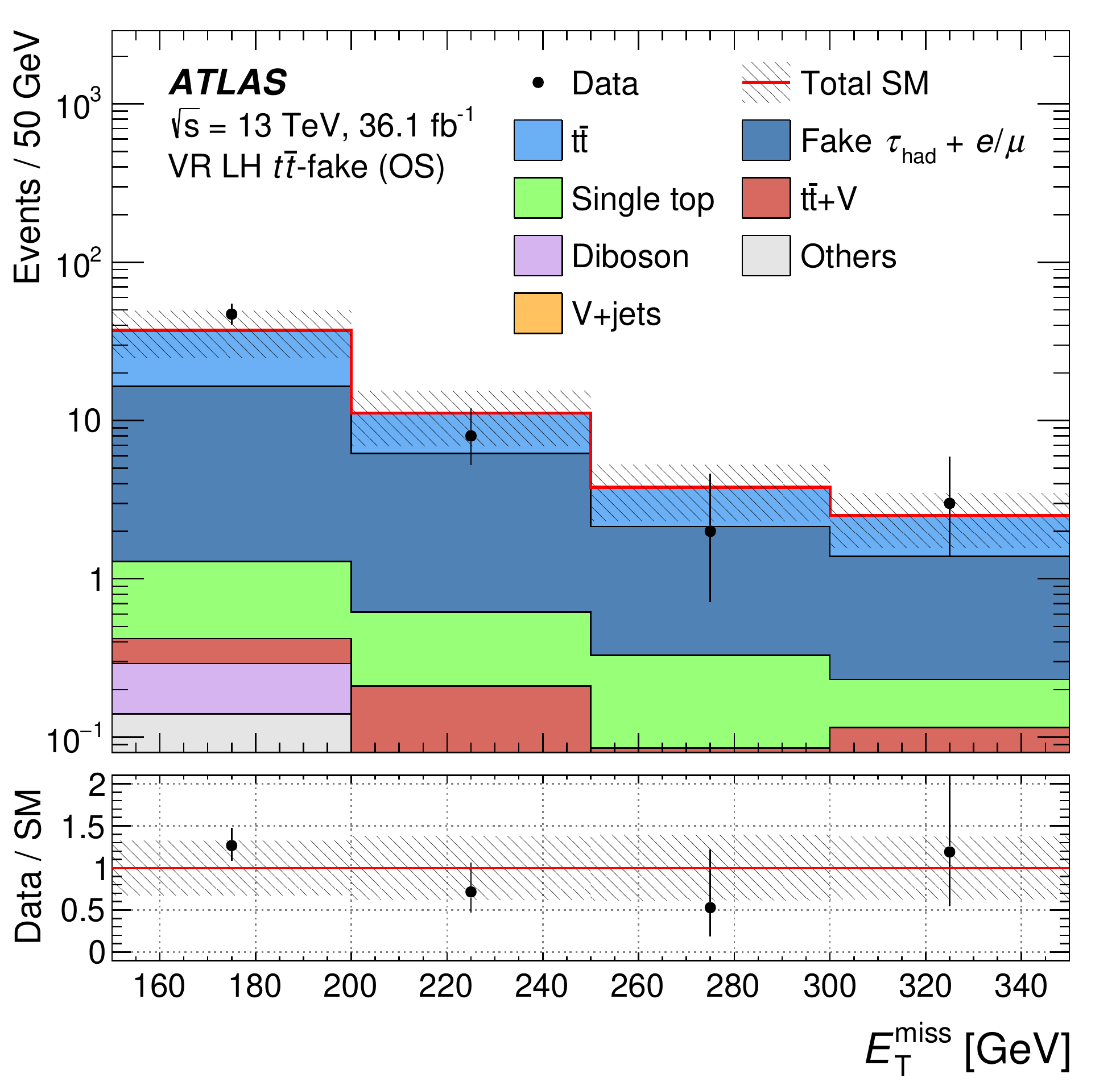} %
  
  \caption{\discussID{1}
    Distributions of $\mttwoLT$ (left) and \met (right)
      in the control region and the validation regions of the lep-had channel,
      \CRLHttReal (top left), 
      \VRLHttReal (top right),
      \VRLHttFakeSS (bottom left), and
      \VRLHttFakeOS (bottom right).
    \VerticalLine{\mttwoLT}{\CRLHttReal} 
    (The range from 60 to \GeV{100} in the top left plot corresponds to \VRLHttReal.)
    \StackedSM
    \BandIsStatSyst
    \FakesFromFFM
    \RightBinOverflow
  }
  \label{fig:CRVR_lephad}
\end{figure}

Distributions of the main discriminating variables \mttwoLT and \met in the CR and the three VRs of the lep-had channel
  are shown in \fig{fig:CRVR_lephad}.
The normalization obtained from the background fit (cf.\ \tab{tab:mus}) %
  is used for \ttbar production with real tau leptons, \ttV and diboson production. 
For single-top production and \Vjets, %
  the theory prediction for the cross section is used.
All contributions from events with fake tau leptons (labeled ``\alllhfakes'' in the legend)
  are estimated using the fake-factor method.
All other processes, which are expected to give only small contributions, 
  are merged into one distribution (``others'').
All selection requirements are applied in all plots, with the exception of the top left plot,
  where the requirement on \mttwoLT is not applied,
  but indicated by a vertical line instead.
The predicted Standard Model background and the observed data are in good agreement. %
The largest differences are found in the top left plot at $\mttwoLT = \GeV{70}$
  and in the first bin in the top right plot of \met.
They correspond to the small excess in \VRLHttReal,
  which is discussed in \sec{sec:results}.

\subsubsection*{Fake-factor method}
\label{sec:fake-factor-method}

The fake-factor method is used to estimate the contribution of events in the lep-had channel
  in which the reconstructed tau lepton is a fake.
This estimate is obtained as the product of the number of events passing a selection 
  based on looser tau identification requirements 
  and the fake factor,
  which relates the number of events with looser tau-lepton candidates 
  to the number where tau leptons meet the nominal identification criteria.

\begin{linenomath*} %
To compute the fake factor $F$, a looser set of criteria for the tau identification is used (``AntiID''), %
  which is orthogonal to the default working point used in the analysis (``ID''), cf.\ \sec{sec:eventreconstruction}.
The value $F$ is the ratio of the number of events with a tau lepton passing the ID requirements to the number passing the AntiID requirements in the measurement region (MR) in data; these numbers are denoted 
  $N^\star(\text{data, MR})$, %
  where $^\star$ is ID or AntiID.
It depends on the \pt %
  and the number of tracks associated with the tau-lepton candidate.
No strong dependence on the pseudorapidity is observed.
As the contribution of electrons misidentified as tau leptons is small compared to that from jets,
  differences in the fake composition 
  between the measurement region and the signal region are not expected to have significant impact on the estimate.
The contamination from events with real tau leptons $N^\star_\text{real}(\text{MC, MR})$ is estimated from simulation 
  and subtracted when taking the ratio,
\begin{equation*}
  F = \frac{N^{\text{ID}}(\text{data, MR}) - N^{\text{ID}}_{\text{real}}(\text{MC, MR}) }
            {N^{\text{AntiID}}(\text{data, MR}) - N^{\text{AntiID}}_{\text{real}}(\text{MC, MR}) }.
  \label{eq:FF_def}
\end{equation*}
The measurement region is chosen such that this contamination is as small as possible.
Overall, the contamination is about \percent1 for AntiID 
  and about \percent{10} for ID tau leptons. %
It is \pt-dependent and increases up to \percent{25} at high \pt for ID tau leptons. %

The number of events with fake tau leptons passing the target selection (TR) %
  is then estimated as
\begin{equation*}
  N_{\text{fakes}}             (\text{TR}) = \left( N^{\text{AntiID}}(\text{data, TR}) - N^{\text{AntiID}}_{\text{real}}(\text{MC, TR}) \right) \cdot F,
  \label{eq:FF_appl}
\end{equation*}
where again $N^{\text{AntiID}}_{\text{real}}(\text{MC, TR})$ is a correction 
  that accounts for the contamination from events with real tau leptons
  and is estimated using simulation.
\end{linenomath*}
Both the number of events with looser tau identification in the target selection
  and the fake factor are obtained from data.
The only inputs taken from simulation are the small corrections 
  that account for events with real tau leptons.

The measurement region in which the fake factors are determined is based on the lep-had preselection.
Events are selected where the tau lepton has the same charge as the light lepton to increase the fraction of fake tau leptons.
The largest contribution to the events with fake tau leptons in the signal region,
  which is estimated with the fake-factor method,
  is from \ttbar production.
Therefore, a requirement of $\met > \GeV{100}$ is applied 
  and at least one $b$-tagged jet required
  to also obtain a high purity in \ttbar~events in the measurement region.
Finally, $\mttwoLT < \GeV{60}$ is required to make the measurement region orthogonal to the same-charge validation region \VRLHttFakeSS.
The fake factors determined in the measurement region vary between 0.22 (0.041) and 0.085 (0.009) for 1-prong (3-prong) tau leptons as a function of \pt.

\subsection{Had-had channel}
\label{sec:CRhadhad}

\begin{table}[ht]
  \centering
  \begin{tabular}{lrrrrrr}
    \toprule
                                &        {CR HH {\ttbar}-fake} & {CR HH {\ttbar}-real}  &&    {VR HH {\ttbar}-fake}  & {VR HH {\ttbar}-real}  & SR HH         \\ 
    \midrule                                                                             
    Charge($\tau_1$,$\tau_2)$   & ---                          & opposite               && ---                       & opposite               & opposite      \\
    $\mttwoTT$                  & $<\GeV{30}$                  & $<\GeV{30}$            && \GeVrange{30}{80}         & \GeVrange{30}{80}      & $>\GeV{80}$   \\
    $\met$                      & $>\GeV{120}$                 & $>\GeV{120}$           && $>\GeV{160}$              & $>\GeV{160}$           & $>\GeV{200}$  \\
    $\mT(\tau_1)$               & $<\GeV{70}$                  & $>\GeV{70}$            && $<\GeV{100}$              & $>\GeV{100}$           & ---           \\
    $m(\tau_1,\tau_2)$          & $>\GeV{70}$                  & $>\GeV{70}$            && ---                       & ---                    & ---           \\
    \bottomrule
  \end{tabular}
  \caption{\discussID{9}
    Definitions of the \ttbar control and validation regions and the signal region in the had-had channel. 
    \TauOneTwo
    \DashNoCut
    \BracketRange
    A common preselection as given in \tab{tab:presel_compare} for the had-had channel is applied.
  }
  \label{tab:def_CRVR_hadhad}
\end{table}
Two control and two validation regions are defined for the background with pair production of a top and an anti-top quark in the had-had channel.
In all of these regions,
  the preselection %
  requirements for the had-had channel from \tab{tab:presel_compare} are applied.

As in the lep-had channel,
  the sequence of control regions, validation regions, and signal region is ordered by increasing \mttwo, the main discriminating variable.
The CRs are restricted to $\mttwo < \GeV{30}$, and the SR requires $\mttwo > \GeV{80}$.
The VRs cover the intermediate phase-space region $\GeV{30} < \mttwo < \GeV{80}$,
  so that the extrapolation in \mttwo from the CRs to the SR can be validated here.
A separation between events with real and fake tau leptons
  is achieved using the transverse mass calculated from the leading tau lepton and the missing transverse momentum.
Events with fake tau leptons dominate at low values of \mT;
  events with real tau leptons tend to have higher values of \mT.
In the signal region, the two tau leptons are required to have opposite charge,
  but since in events with a fake tau lepton the relative sign of the electric charges of the tau leptons is random, %
  the number of events with fake tau leptons in the fake CR and VRs is increased
  by not imposing this requirement.
Also, the requirement on \met is lowered to \GeV{120} to increase the number of events in the CRs.
A requirement on the invariant mass of the tau-lepton pair suppresses \Zjets events and increases the purity in \ttbar~events in the CRs.
\Tab{tab:def_CRVR_hadhad} summarizes the definitions of the CRs and VRs in the had-had channel.

\begin{figure}
  \centering
  \includegraphics[width=0.5\textwidth]{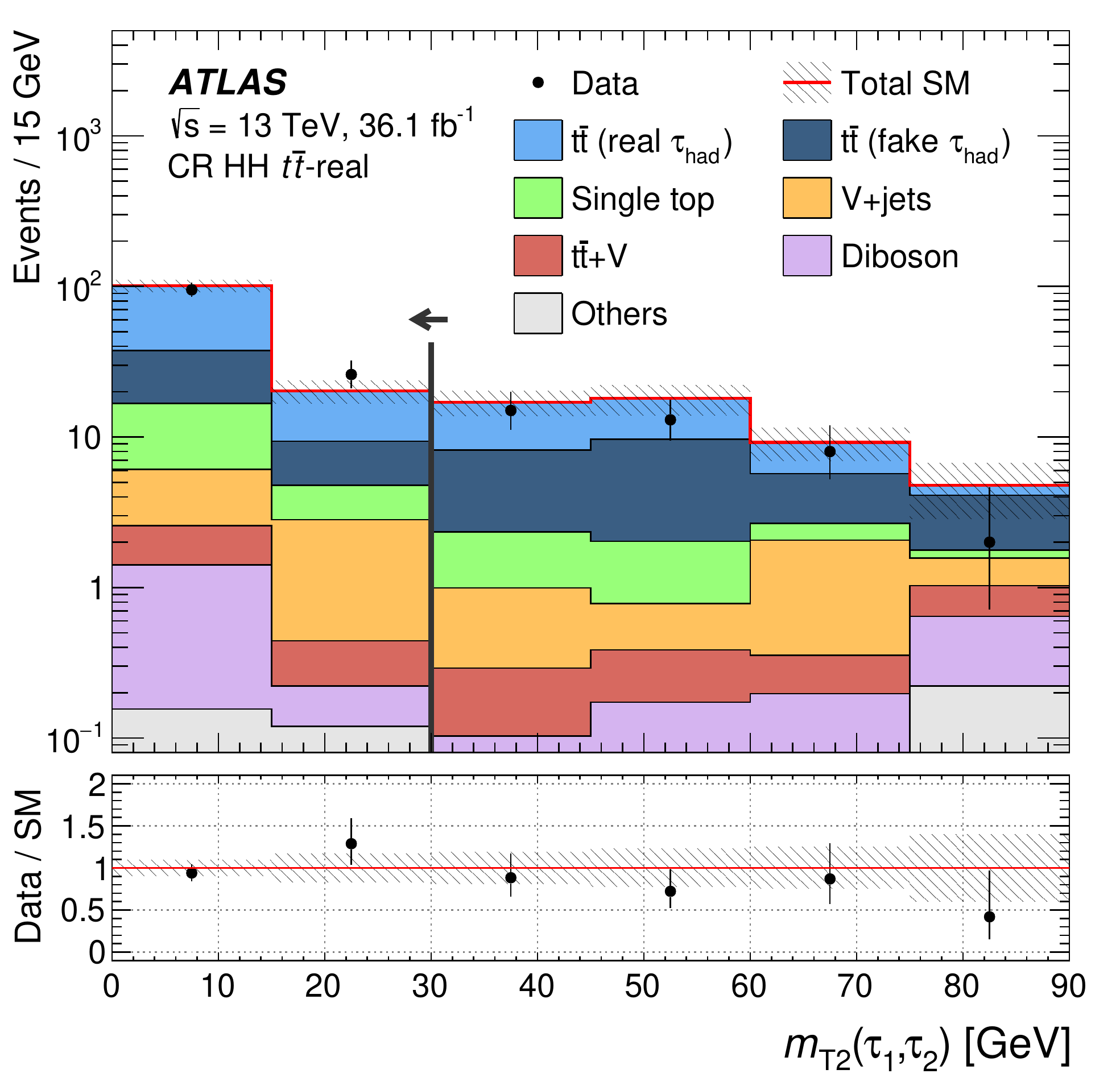}%
  \includegraphics[width=0.5\textwidth]{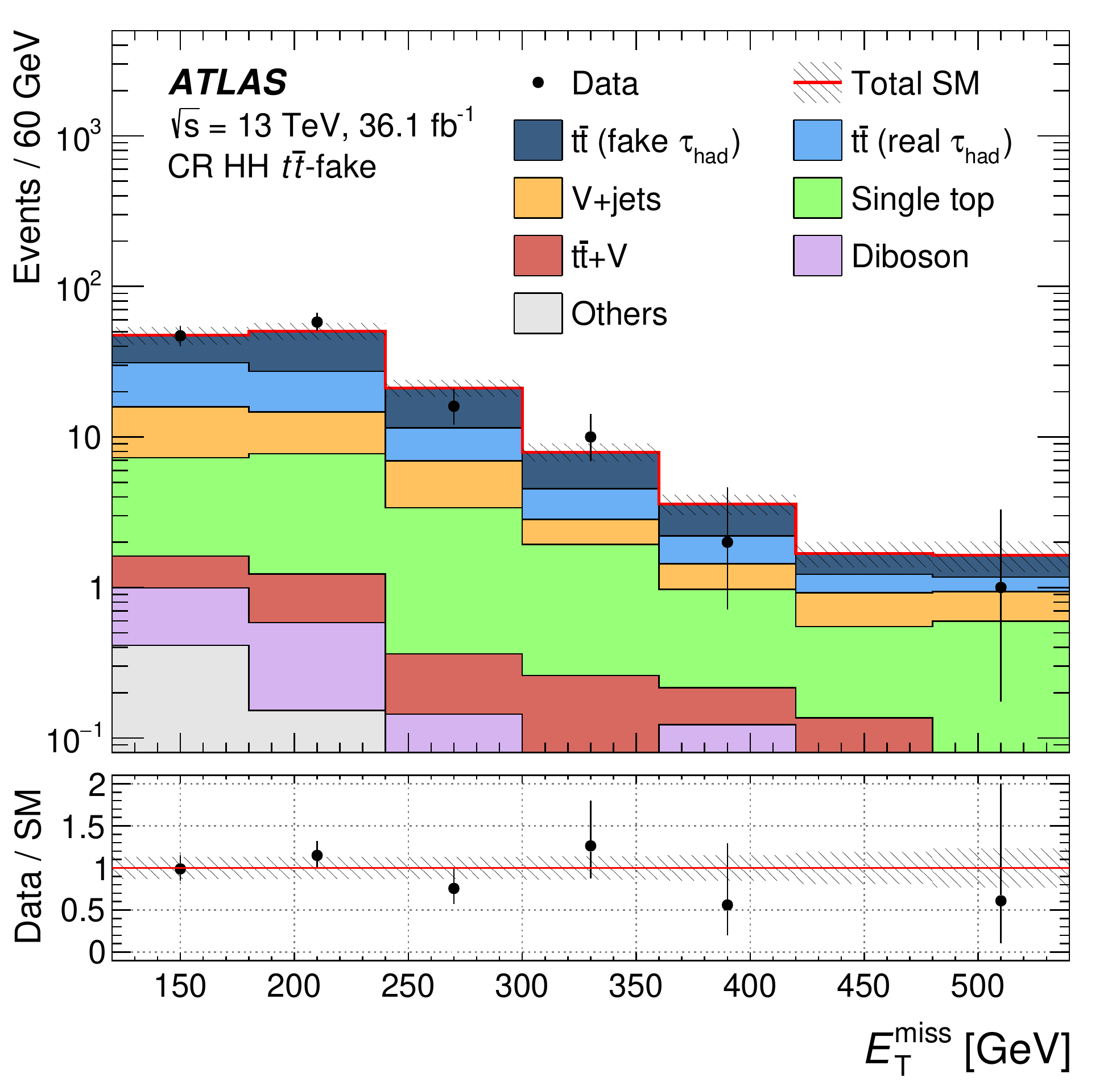}
  
  \includegraphics[width=0.5\textwidth]{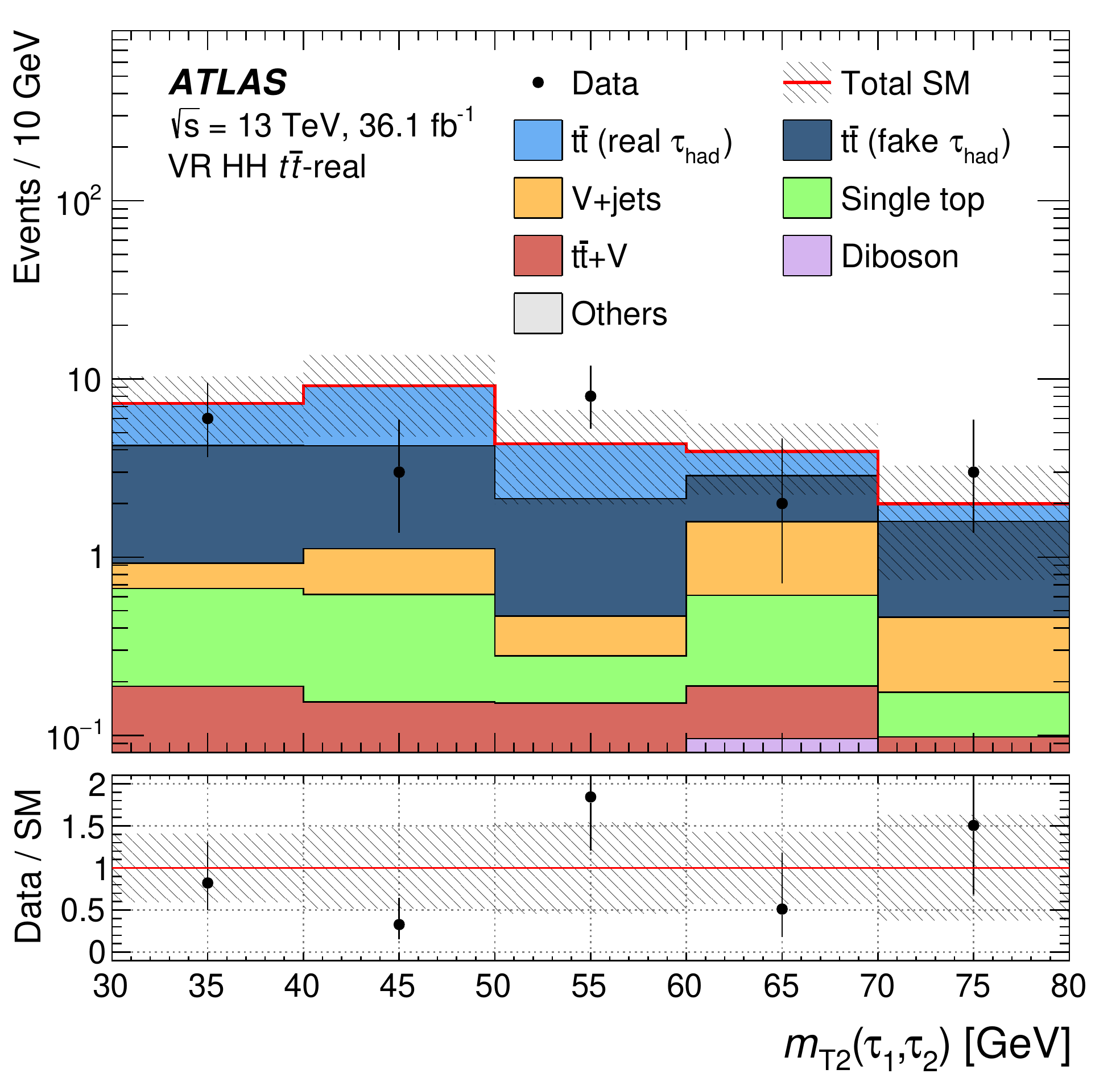}%
  \includegraphics[width=0.5\textwidth]{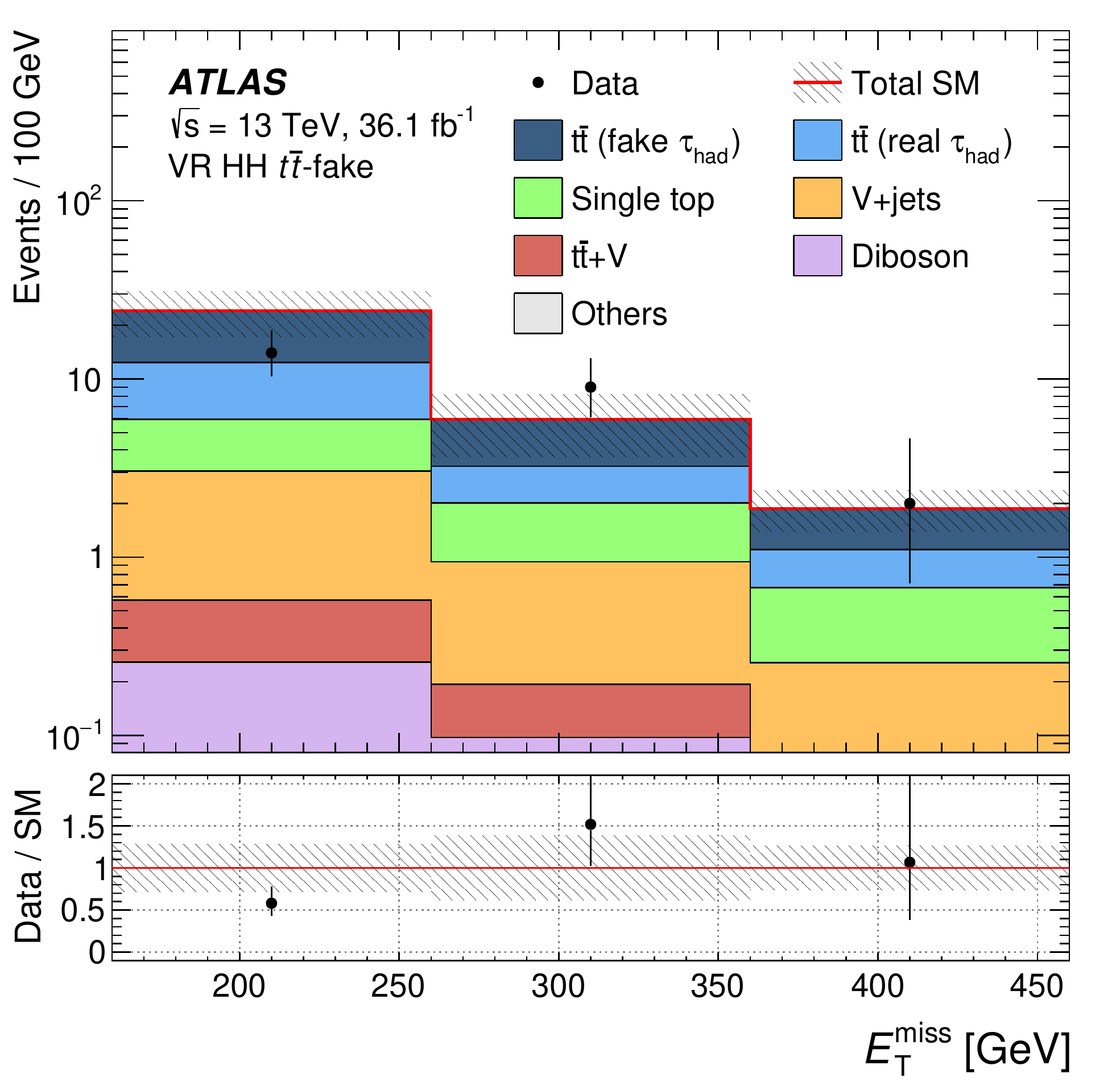}
  \caption{\discussID{2}
    Distributions of \mttwoTT (left) and \met (right)
      in two control and two validation regions in the had-had channel,
      CR HH {\ttbar}-real (top left),
      CR HH {\ttbar}-fake (top right),
      VR HH {\ttbar}-real (bottom left), and
      VR HH {\ttbar}-fake (bottom right).
    \TauOneTwo
    \VerticalLine{\mttwoTT}{CR HH {\ttbar}-real}
    \StackedSM
    \BandIsStatSyst
    \RightBinOverflow
    In the lower left plot,
      the overflow contribution is zero
      because VR HH {\ttbar}-real has an upper bound on \mttwo.
  }
  \label{fig:CRVR_hadhad}
\end{figure}

Distributions of the main discriminating variables \mttwoTT and \met in the two CRs and two VRs of the had-had channel
are shown in \fig{fig:CRVR_hadhad}.
The simulation-based estimates for \ttbar production, separated into real and fake tau-lepton contributions, and for \ttV and diboson production
  are scaled with the normalization factors obtained from the background fit (cf.\ \tab{tab:mus}).
The background process ``\ttbarfakes'' includes both the events with one real and one fake tau lepton and 
events with two fake tau leptons.
The purity ranges between \percent{41} and \percent{61} in the four control and validation regions.

The relative contributions of events selected by each of the two triggers used in the had-had channel (cf.\ \sec{sec:presel})
  vary between the control and validation regions and the signal region,
  as the fraction of events selected by the \met trigger becomes higher with an increasing \met requirement.
The normalization factors were therefore recomputed for the two sets of events selected exclusively by one of the two triggers.
They are compatible within their statistical uncertainties,
  showing that there is no dependence of the normalization factors on the trigger selection.
This is also confirmed by good agreement between data and predicted background yields in the validation regions
  when the normalization factors derived in the control regions are applied.

\subsection{Common control regions}
\label{sec:CRcommon}

%
%
%
%
%

\begin{table}[h]
  \centering
  \begin{tabular}{lrr}
    \toprule                                                                                          
                                       & CR \ttV                 & CR \VV                             \\
    \midrule                                                                                          
    $\pT({\text{jet}}_{2})$            & $>\GeV{26}$             & $>\GeV{26}$                        \\
    $n_\text{SFOS}$                    & $\geq 1$                & $\geq 1$                           \\ 
    \mZclosest                         & \GeVrange{80}{100}      & \GeVrange{80}{100}                 \\
    $n_{\bjet}$                        & $ \geq 2$               & $0$                                \\
    $n_\text{lepton}$                  & $ \geq 3$               & $ \geq 2$                          \\
    $n_\text{lepton} + n_\text{jet}$   & $ \geq 6$               & ---                                \\
    $\met/\sqrt{H_{\text T}}$          & ---                     & $> \unit[15]{\sqrt{\text{GeV}}}$   \\
    $\mttwo(\ell,\ell)$                & ---                     & $>\GeV{120}$                       \\
    \bottomrule
  \end{tabular}
  \caption{\discussID{11}
    Definition of the \ttV and \VV control regions.
    The total number of signal leptons ($e$, $\mu$ or $\tauh$) is given by $n_\text{lepton}$,
      and $n_\text{SFOS}$ is the number of lepton pairs with the same flavor and opposite charge.
      Other variables are defined in the text.
    \DashNoCut
    \BracketRange
  }
  \label{tab:def_CR_common}
\end{table}
The definitions of the CR for events with \ttbar production in association with a vector boson, CR \ttV,
  and of the CR for events with diboson processes, CR \VV, 
  are given in \tab{tab:def_CR_common}.
They do not use the common preselection described in \sec{sec:presel}
  but select events with at least two signal leptons ($e$, $\mu$ or $\tauh$). %
These events also need to have fired the single-lepton trigger
  and the respective trigger plateau requirement is applied as described in \sec{sec:presel}, %
  so that at least one light lepton must be among the two leptons.
Two jets must be present with $\pt > \GeV{26}$.
No $b$-tagged jets are allowed in CR \VV,
  whereas in CR \ttV
  at least two $b$-tagged jets are required to select events with top-quark decays.

The \ttV background in the signal region mostly consists of events in which a \ttbar pair is produced in association with a $Z$ boson
  that decays to two neutrinos providing large \met.
This type of background cannot easily be separated from other backgrounds,
  in particular pure \ttbar production,
  so that instead a CR enriched in $\ttZ$ with $Z\to\ell\ell$ is used.
It is then assumed that the normalization factor derived for this process is also valid for the $Z$ boson decaying to neutrinos.
Furthermore, as events with four or more leptons are too rare to make a CR,
  the CR \ttV also accepts events with only one additional, third signal lepton.

To select events with $Z$-boson decays,
  the invariant mass of each same-flavor, opposite-charge (SFOS) lepton pair in the event is calculated.  
The pair with invariant mass closest to the mass of the $Z$ boson 
  is selected and assumed to originate from the $Z$-boson decay.
The invariant mass of this pair, \mZclosest, is required to be within about \GeV{10} of the $Z$-boson mass.
As the invariant mass computed from the visible products of a $Z$-boson decay to hadronically decaying tau leptons
  is smaller than the $Z$-boson mass, this in effect removes most of the events with tau-lepton pairs. %
After applying these requirements, there is still a sizable contribution from \Zjets events, %
  where the SFOS pair originates from the $Z$ boson and one of the jets is misidentified as a tau lepton.
Requiring the total number of leptons and jets to be at least six
  gives a small increase in the purity in \ttZ events in this region. 

Events with diboson production entering the signal regions
  mostly have either two or three charged leptons.
Events with four leptons are negligible in both channels.
A CR for diboson production based on a pure tau-lepton selection
  would suffer from a high contamination from events 
  in which a $W$ boson is produced in association with jets,
  one of which is misidentified as a hadronically decaying tau lepton.
Therefore, the CR selection includes all lepton flavors
  and makes use of \mttwo and the significance of the \met, measured as $\met/\sqrt{\HT}$,
  to suppress \Zjets events. %
The requirement on \mZclosest is used %
  to suppress signal contamination, %
  which otherwise becomes non-negligible 
  for small mass differences between the top squark and tau slepton in the simplified model.
The composition of different diboson processes in the signal region is similar to that of the control region.
\Fig{fig:CR_common} shows the distribution of \met in CR \ttV and %
  in CR \VV
  with the normalization factors from the background fit (cf.\ \tab{tab:mus}) applied.
The purity is about \percent{79} in CR \ttV and \percent{91} in CR \VV.

\begin{figure}
  \centering
  \includegraphics[width=0.5\textwidth]{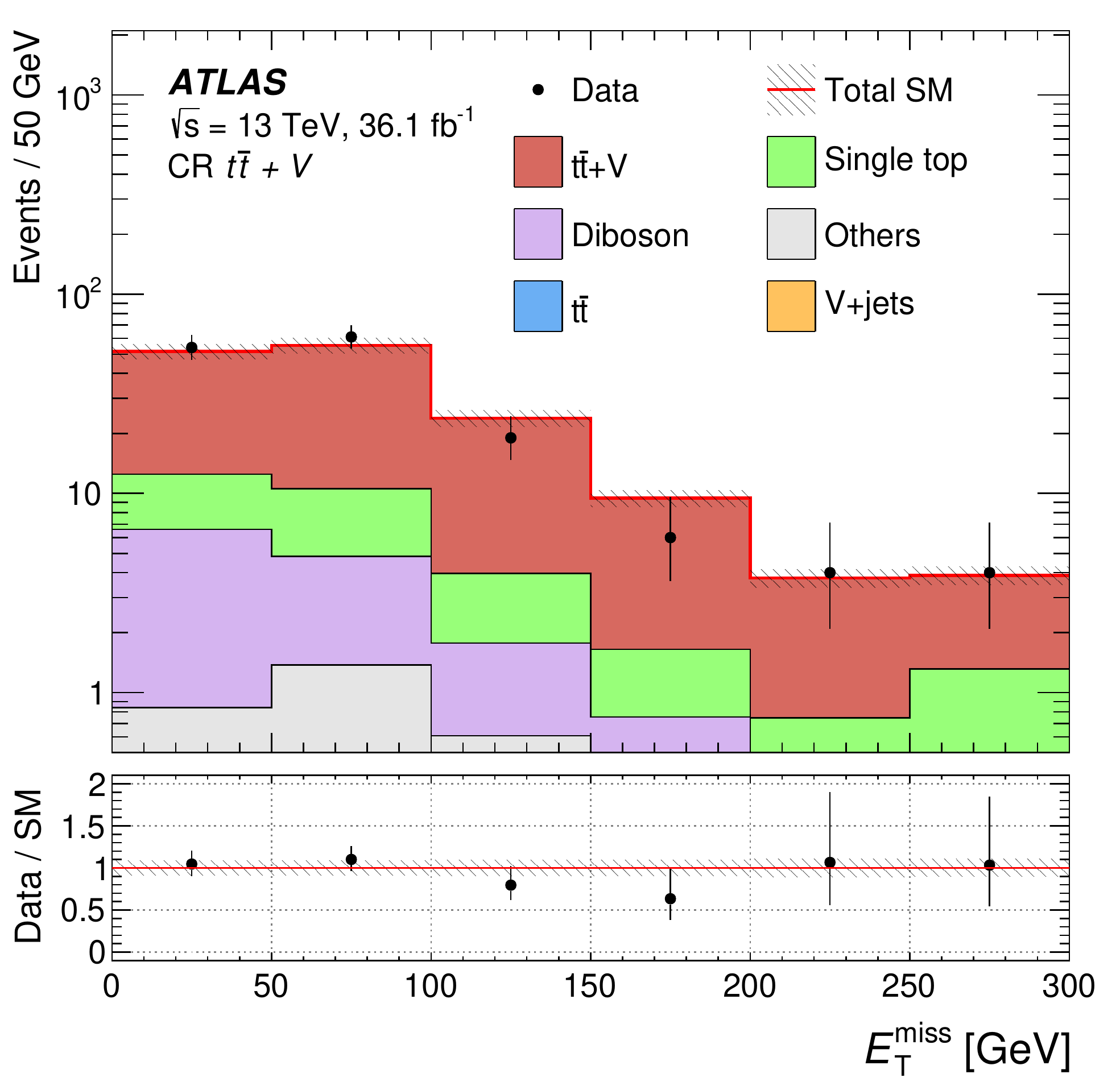}%
  \includegraphics[width=0.5\textwidth]{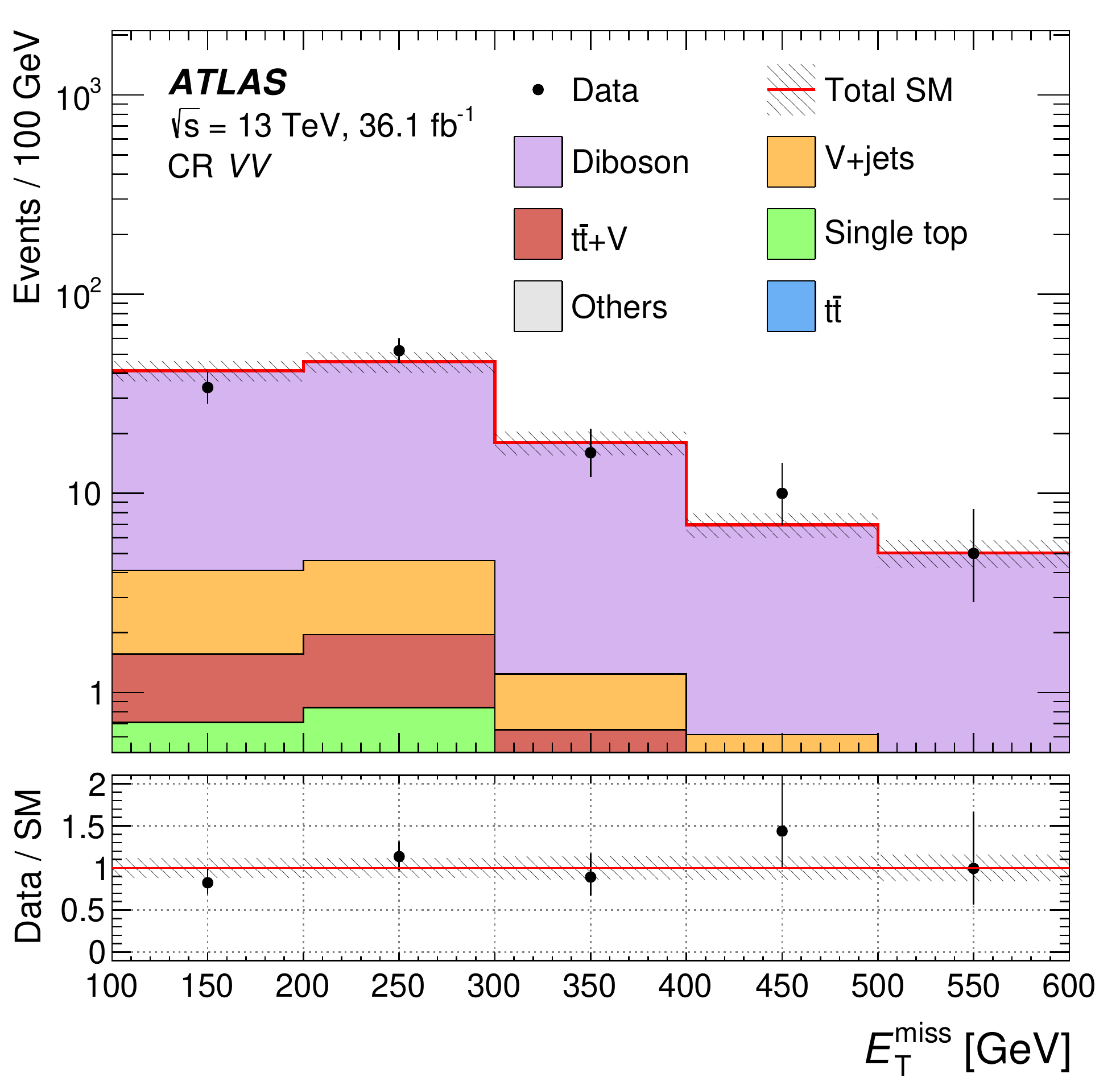} %
  \caption{\discussID{16}
    Distributions of \met in CR \ttV (left) and CR \VV (right).
    \BandIsStatSyst
    \RightBinOverflow
    The lower bins of \met in CR \VV which are not shown in the right plot 
      are empty due to the requirement on $\met/\sqrt{\HT}$.
  }
  \label{fig:CR_common}
\end{figure}

\section{Systematic uncertainties}
\label{sec:systematics}

Experimental systematic uncertainties are taken into account for all simulated background and signal samples.
For leptons, experimental systematic uncertainties arise from the reconstruction and identification efficiencies, %
  and for electrons and muons also from the isolation efficiency. %
For jets, additional uncertainties from the pileup subtraction, pseudorapidity intercalibration, flavor composition, and punch-through effects, %
  as well as uncertainties in the flavor-tagging and jet-vertex tagging efficiencies are considered %
  using a reduced set of nuisance parameters~\cite{ATL-PHYS-PUB-2015-015}.
Uncertainties in the energy resolution and calibration are taken into account for all physics objects. %
The \MET has an additional uncertainty due to the contribution of the soft-track term. %
The fast detector simulation used for the signal samples
  brings additional uncertainties in jets and tau leptons. %
Further sources of experimental systematic uncertainty are the pileup reweighting of simulated events
  to cover the uncertainty in the ratio of the predicted and measured inelastic cross sections, %
  and the measurement of the trigger scale factors. %

Several sources of uncertainty are found to be important for the background estimate obtained from the fake-factor method.
Statistical uncertainties in the fake factors 
  from the number of events in the measurement region 
  and the number of AntiID events in the respective target selection %
  are propagated into the uncertainty in the final estimate.
Further uncertainties in the fake factors arise from the contribution of multi-jet events,
  which enter the measurement region due to the softer requirement on \met relative to the other lep-had selections,
  and the subtraction of events with real tau leptons.
The former uncertainty is estimated by varying the \met requirement of the measurement region, 
  and the latter by scaling the simulation-based estimate for these events by up to \percent{$\pm40$}.
An uncertainty from the choice of AntiID working point is derived by reevaluating %
  and comparing 
  the estimates obtained from the fake-factor method for different values of the AntiID working point.
Finally, the impact of the extrapolation of the fake factor in \mttwo is translated into an uncertainty
  by comparing fake factors obtained for different ranges of \mttwo in the measurement region.
This is the dominant source of uncertainty in the fake-factor method.

Uncertainties in the theoretical modeling are evaluated for the dominant processes selected in the analysis.
For the hard-scatter modeling of the \ttbar and single-top processes, 
  systematic uncertainties are estimated 
  by comparing the hard-process generation between \powheg and \MGMCatNLO, 
  both interfaced to \herwigpp for the parton showering.
Uncertainties in the fragmentation and hadronization are estimated 
  from a comparison of samples generated with \powheg for the hard scattering 
  interfaced to \herwigpp or \pythia for the parton shower. %
Uncertainties in additional radiation are obtained through a variation of the generator settings, 
  such as those for the produced shower radiation, the factorization and renormalization scales and the NLO radiation. %
An uncertainty in the treatment of the interference subtraction of single-top-quark production in the $Wt$ channel and \ttbar production at next-to-leading order
  is estimated as the difference between diagram-removal and diagram-subtraction schemes~\cite{ATL-PHYS-PUB-2016-004,Frixione:2008yi}. %

For \ttV production, the uncertainty in the hard-scatter, fragmentation and hadronization modeling is assessed
  by comparing the nominal \MGMCatNLO samples interfaced to \pythia %
    to samples generated with \sherpa.
For diboson and \Vjets production, the nominal \sherpa samples are compared 
  to samples generated with \powheg or \MGMCatNLO,
    both interfaced to \pythia for the parton showering. %
For \ttV, \VV, and \Vjets, the effects of additional variations of the internal parameters of the generators
  for the factorization and hadronization scales are evaluated. %

An additional cross-section uncertainty of \percent{5} is considered for \Zjets, \Wjets~\cite{Aad:2016naf},
  and single-top-quark production~\cite{Kidonakis:2010ux,Kidonakis:2010tc,Kidonakis:2011wy}
  because their yields are not normalized in control regions.
The uncertainty in the integrated luminosity described in \sec{sec:dataset}
  is also applied to all backgrounds that are taken directly from simulation.
In all regions, %
  the statistical uncertainties in the MC simulations  
  and the uncertainties in the normalization factors are taken into account. %

\begin{table}
\begin{center}

\begin{tabular}{lcc}
\toprule
                                         & SR LH                  & SR HH                  \\
\midrule
Total systematic uncertainty             & $\pm\, \percent{29}$ & $\pm\, \percent{53}$ \\
\midrule
Fake-factor method                       & $\pm\, \percent{23} $  & ---                      \\
Jet-related                              & $\pm\, \percent{9.4} $   & $\pm\, \percent{36} $  \\
Tau-related                              & $\pm\, \percent{7.2} $   & $\pm\, \percent{32} $  \\
Other experimental                       & $\pm\, \percent{6.2} $   & $\pm\, \percent{12} $  \\   [0.3mm]
Theory modelling                         & $\pm\, \percent{8.4} $   & $\pm\, \percent{20} $  \\
MC statistics                            & $\pm\, \percent{7.5} $   & $\pm\, \percent{17} $  \\
Normalization factors                    & $\pm\, \percent{4.8} $   & $\pm\, \percent{14} $  \\
Luminosity                               & $\pm\, \percent{0.3} $  & $\pm\, \percent{0.8} $  \\

\bottomrule
\end{tabular}

\end{center}
\caption{
  Relative systematic uncertainties in the estimated number of background events in the signal regions (left: lep-had, right: had-had channel).
  In the lower part of the table, a breakdown into different categories is given:
    all jet- and tau-related systematics are added into a respective combined value, 
    while the smaller experimental uncertainties from electrons, muons, flavor-tagging, \MET, and pileup reweighting 
    are combined into ``Other experimental''.
  The percentage values give the relative post-fit uncertainties in the total expected background yield.
  The individual contributions do not add up to the total given in the first row 
    due to the correlations between the individual systematic uncertainties.
}
\label{tab:systs_SR}
\end{table}

The full set of systematic uncertainties in the total background yields is summarized in \tab{tab:systs_SR}.
The largest sources of experimental systematic uncertainty in both channels
  include the jet and tau energy calibration, the pileup reweighting and the \MET measurement. %
In the lep-had channel, the dominant contribution to the overall systematic uncertainty comes from the uncertainties in the fake-factor method.
The advantage of using a data-driven method for the largest part of the background is the moderate total uncertainty in this channel 
  compared to the had-had channel, where simulation is used to extrapolate from the control region.
In the had-had channel, the uncertainty in the total background estimate is driven by the uncertainty in the estimate of \ttbar events with fake tau leptons,
  the largest background contribution.
The dominant effects arise from the systematic uncertainty in the tau energy scale %
  and from jet mismodeling due to the simulation-based residual pileup correction, %
  which significantly affect the extrapolation from the control to the signal region.

For the signal, in addition to the experimental uncertainties, %
  theoretical uncertainties in the cross sections %
  are taken from an envelope of cross-section predictions using different PDF sets and factorization and renormalization scales, 
  as described in Ref.~\cite{Borschensky:2014cia}.
They vary between \percent{13} and \percent{20},
  which is similar to the size of the experimental uncertainties in the signal. %

\section{Results}
\label{sec:results}
\begin{table}[t]
  \centering
  \newcommand{\SRyieldsempty}{\multicolumn{4}{c}{~~---}}
  \begin{tabular}{l r@{}l @{\,$\pm\,$} r@{}l  r@{}l @{\,$\pm\,$} r@{}l} %

     \toprule                         
                        & \multicolumn{4}{c}{SR LH} & \multicolumn{4}{c}{SR HH}                     \\
     \midrule                                               
     Observed events    & \multicolumn{4}{c}{~~3}   & \multicolumn{4}{c}{~~2}                       \\
     \midrule                                             
     Total background   & 2&.2                     & 0&.6            & 1&.9   & 1&.0            \\
     \midrule                           
     \Alllhfakes        & 1&.4                     & 0&.5            & \SRyieldsempty             \\ 
     \ttbarfakes        & \SRyieldsempty            & 0&.6            & $^{0}_{0}$&$^{.7}_{.6}$             \\ [+0.2em]    
     \ttbarreals        & 0&.22                    & 0&.12           & 0&.28  & $^{0}_{0}$&$^{.30}_{.28}$ \\ [+0.1em]
     \ttV               & 0&.25                    & 0&.14           & 0&.26  & 0&.12           \\   
     Diboson            & 0&.15                    & 0&.11           & 0&.28  & 0&.13           \\ [+0.0em]
     Single-top         & 0&.10                    & $^{0}_{0}$&$^{.24}_{.10}$ & 0&.13  & 0&.11           \\ [+0.0em]
     \Vjets             & 0&.032                   & 0&.014          & 0&.26   & 0&.09           \\    
     Others             & 0&.082                   & 0&.022          & 0&.09  & 0&.04           \\     
     \midrule                           
     Signal             & 3&.3                    & 0&.7            & 4&.7    & 1&.2              \\ %
     \multicolumn{6}{l}{($\massof{\tone}=\GeV{1100}$, $\massof{\stauone} = \GeV{590}$)} \\
     \bottomrule                           
  \end{tabular}
  \caption{\discussID{15}
    Expected numbers of events from the SM background processes from the background fit 
      and observed event yield in data for the signal regions in the lep-had and had-had channel,
      given for an integrated luminosity of \intlumi.
    The expected yield for the signal model with $\massof{\tone}=\GeV{1100}$ and $\massof{\stauone}=\GeV{590}$ is shown for comparison.
    The uncertainties include both the statistical and systematic uncertainties
      and are truncated at zero.
    \FakesFromFFM
  }
  \label{tab:SR_yields}
\end{table}
\begin{table}
  \begin{center}
  \begin{tabular}{l r@{}l@{}l@{}l}
    \toprule
      Process &  \multicolumn{4}{c}{Normalization factor} \\
    \midrule

                Diboson            &\hspace{2.5em}  1&.&0& $_{-0.3} ^{+0.6}$ \\ [0.4em] %
                 \ttV              & 1&.&39& $_{-0.23} ^{+0.23}$ \\ [0.4em] %
      \ttbarfakes                  &  1&.&2& $_{-0.4} ^{+0.4}$ \\ [0.4em] %
      \ttbarreals                  &  0&.&81& $_{-0.20} ^{+0.20}$ \\ %

    \bottomrule
    \end{tabular}
  \end{center}
  \caption{\discussID{4}
    Normalization factors obtained from the background-only fit.
    The normalization factor for \ttbar events with fake tau leptons is only relevant for the had-had channel.
  }
  \label{tab:mus}
\end{table}

The statistical interpretation of the results is performed using the HistFitter framework~\cite{arxiv:HistFitter}
  that carries out the fitting procedure based on a maximum-likelihood approach 
  and the hypothesis tests utilizing the profile-likelihood ratio as a test statistic with asymptotic formulae~\cite{Cowan2011}. 
All regions are treated as single bins in the likelihood fits,
  \ie no shape information is used.
Systematic uncertainties are implemented as nuisance parameters,
  taking into account potential correlations.
The background fit uses the three CRs of the lep-had and the had-had channels and the two common CRs simultaneously. %
The normalization factors from the background fit are extrapolated to the VRs and SRs in order to obtain the background estimates in these regions, 
  again accounting for correlations between systematic uncertainties.

The results from the background fit 
  for the individual expected contributions of the SM processes and for their sum in the two signal regions are shown in \tab{tab:SR_yields},
  together with the observed yields from the analysis dataset with an integrated luminosity of \intlumi.
\Tab{tab:mus} summarizes the four normalization factors obtained from the background fit. 
Overall, they are compatible with unity. %
The observed data yields in the signal regions in \tab{tab:SR_yields} 
  are in agreement with the expected total background yields from SM processes 
  in both the lep-had and the had-had channels. %
No significant excess is observed.

\begin{figure}[htbp]
  \centering
  \includegraphics[width=0.5\textwidth]{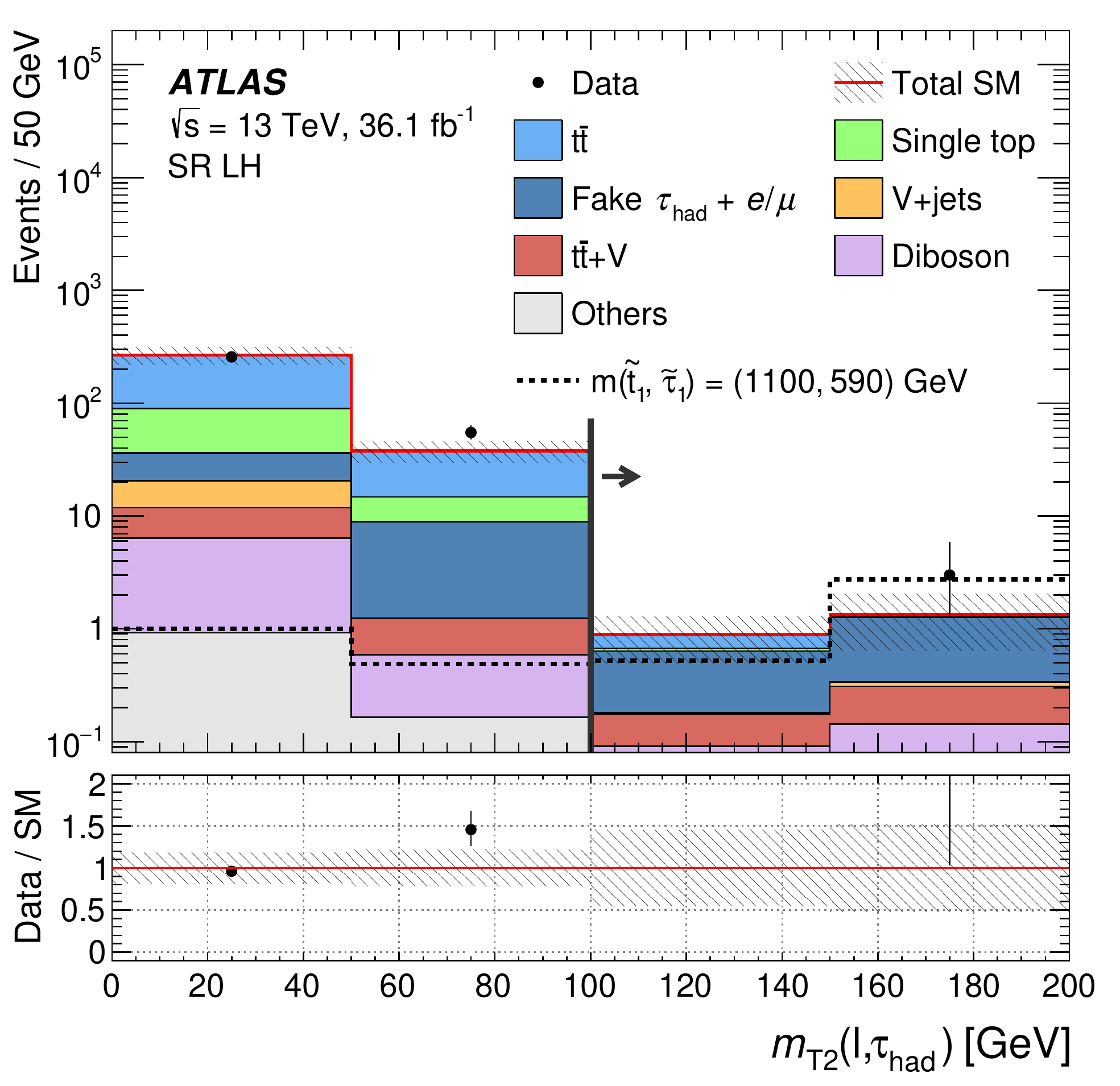}%
  \includegraphics[width=0.5\textwidth]{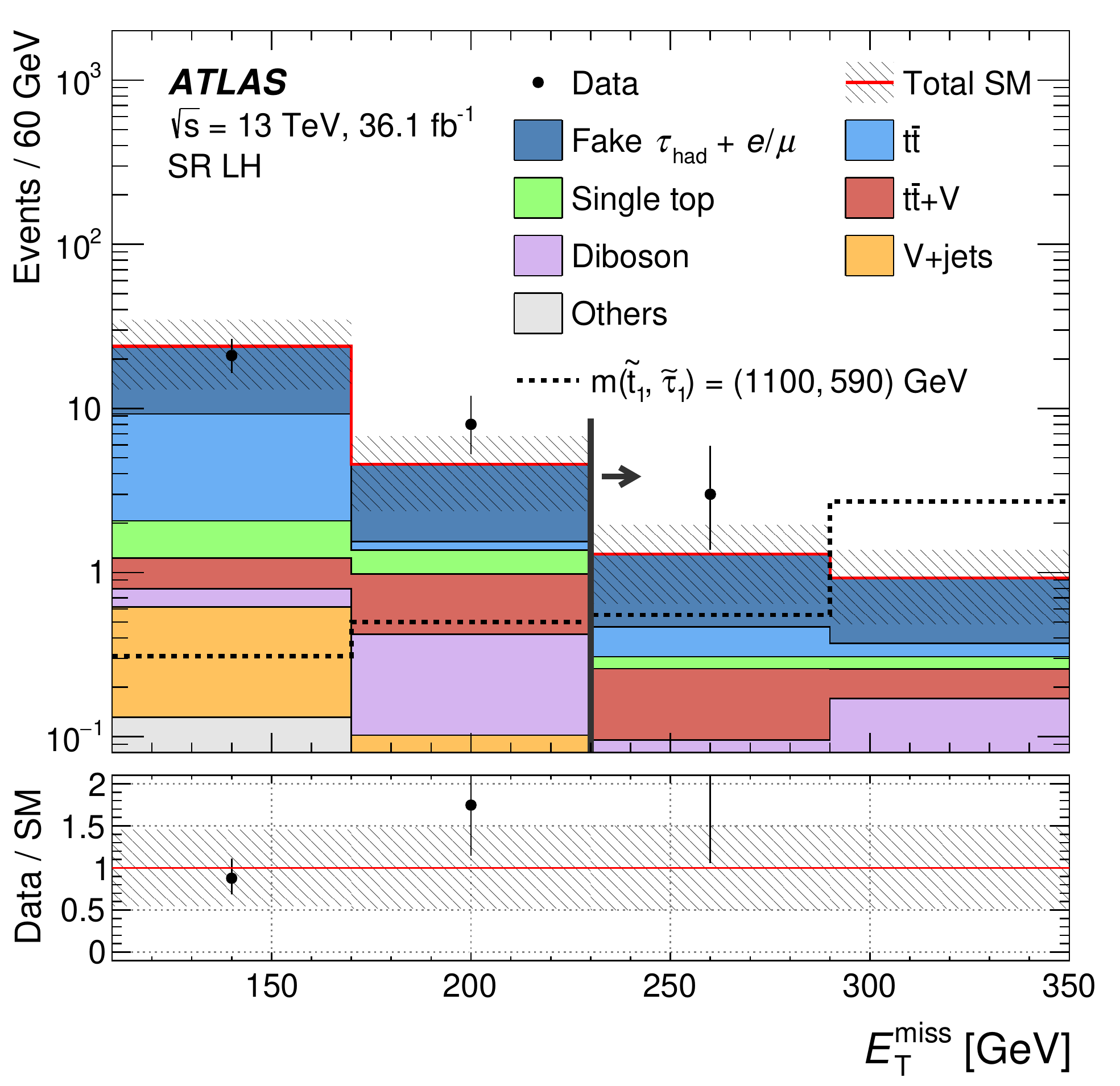}
  
  \includegraphics[width=0.5\textwidth]{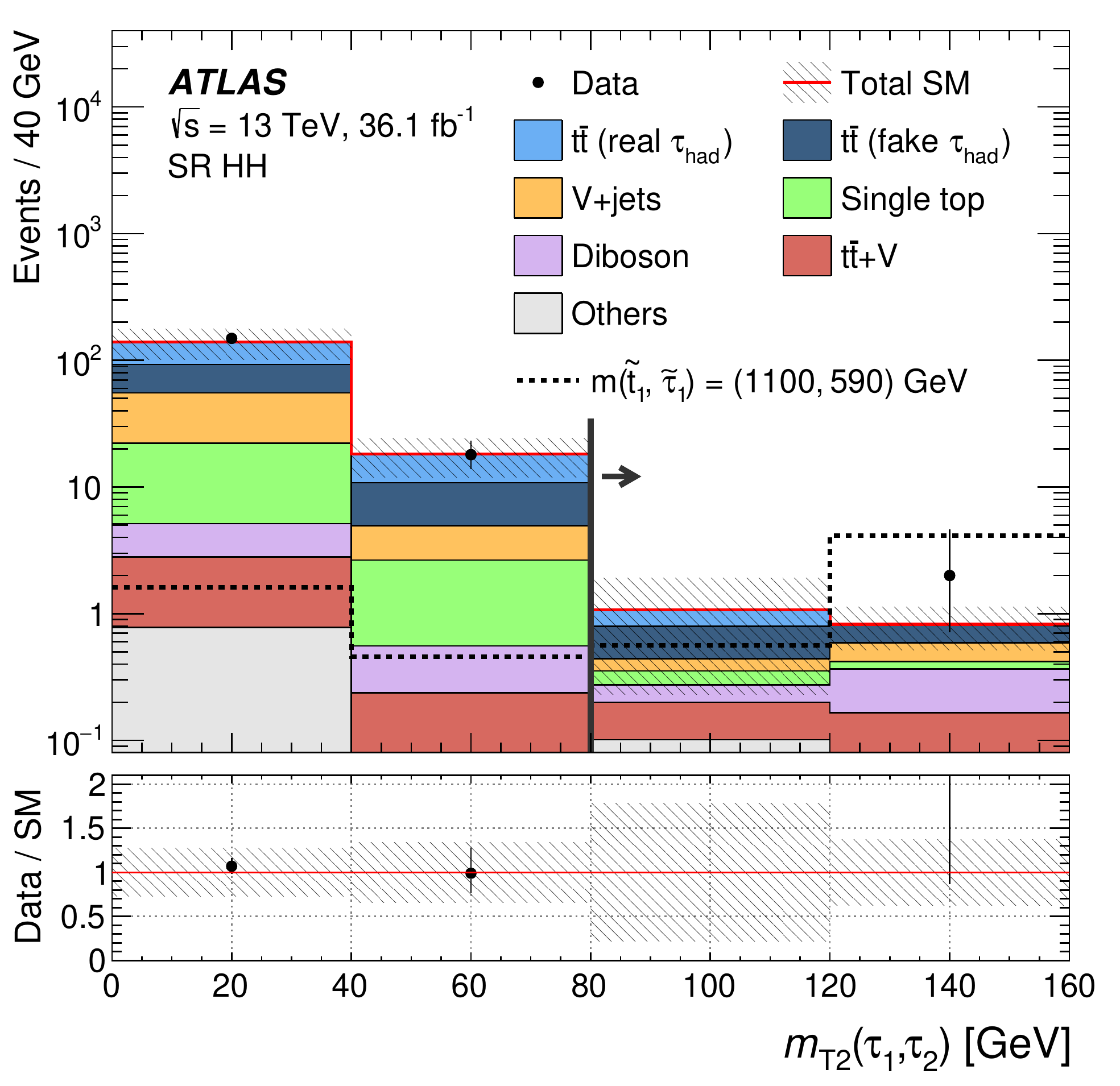}%
  \includegraphics[width=0.5\textwidth]{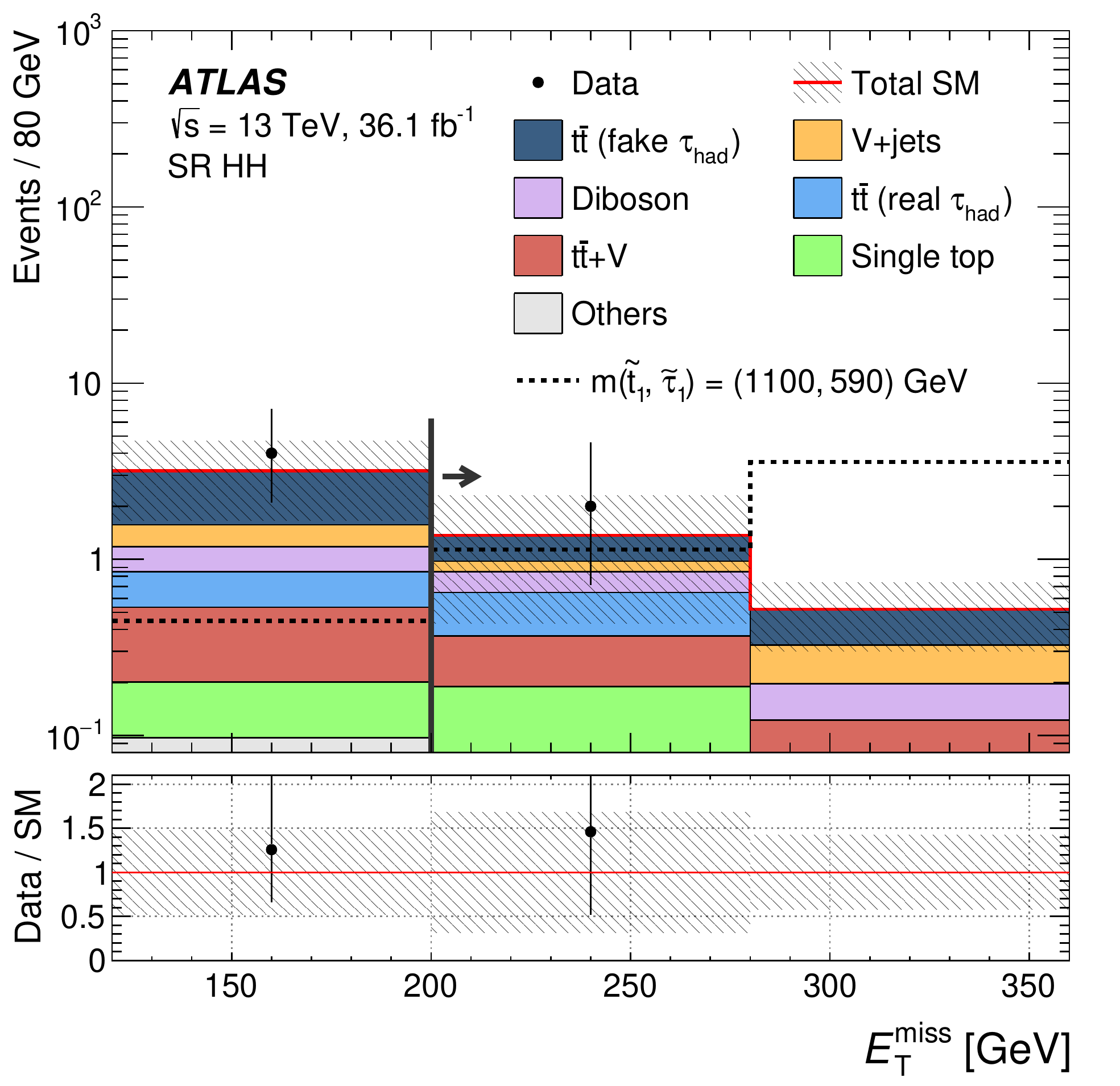}
  \caption{\discussID{5}
    Distributions of \mttwo (left) and \met (right) 
      in the signal regions of the lep-had channel (top) and had-had channel (bottom)
      before the respective selection requirements, indicated by the vertical line and arrow, are applied.
    \TauOneTwo
    \StackedSM
    \FakesFromFFM
    \BandIsStatSyst
    The error bars on the black data points represent the statistical uncertainty in the data yields.
    \DashedSignal
    \RightBinOverflow
  }
  \label{fig:SR_distributions}
\end{figure}

\Fig{fig:SR_distributions} shows the distributions of \mttwo and \met  
  in the signal regions of the lep-had channel and had-had channel.
All selection requirements are applied,
  except that on the variable shown in the plot,
  which is instead indicated by the vertical line and arrow.

\begin{figure}[tbp]
  \centering
  \includegraphics[width=0.95\textwidth,clip,trim=0 25 0 15]{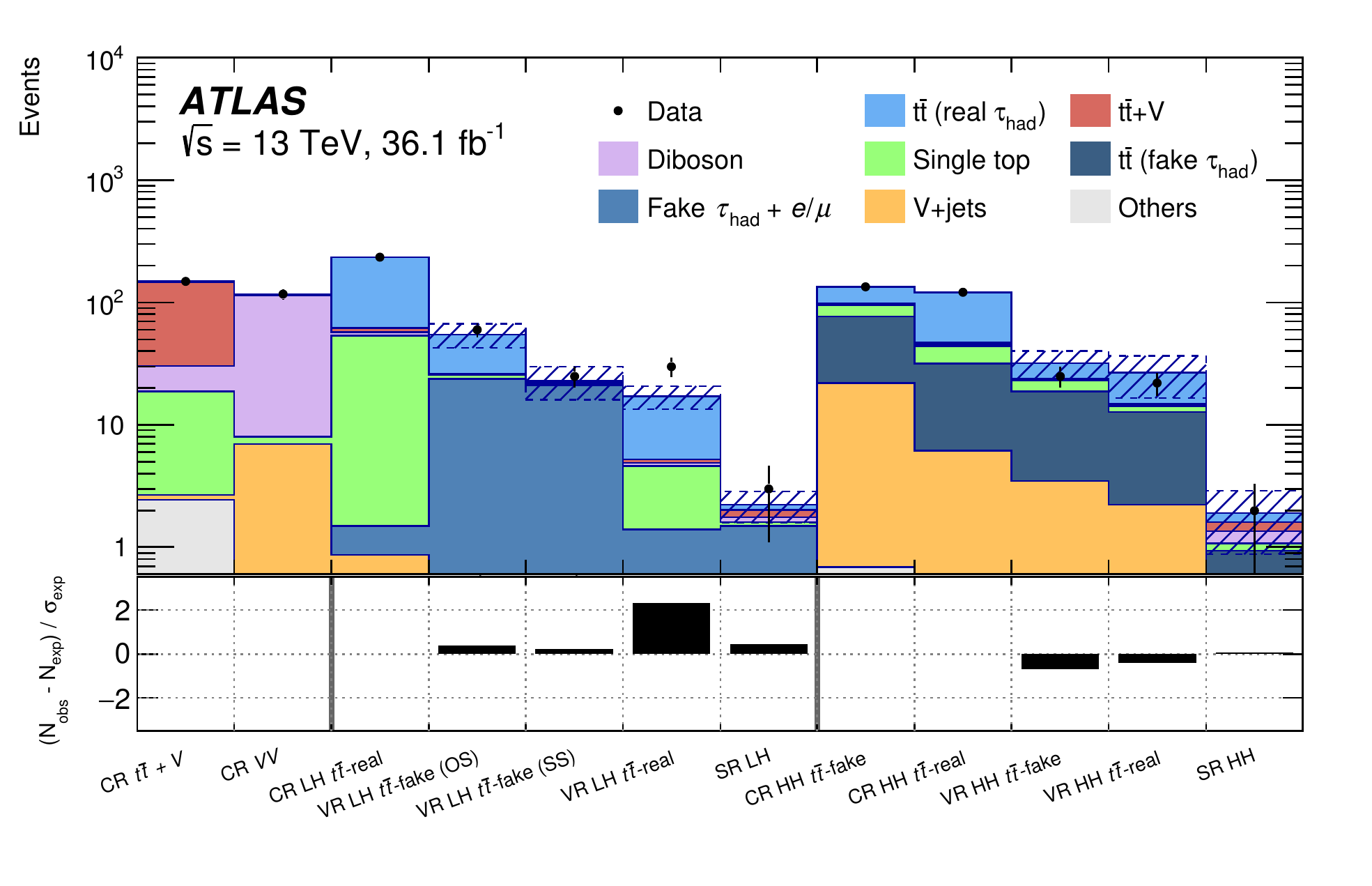} %
  \caption{\discussID{6}
    Data yields and background expectation (top panel) and the resulting pulls (bottom panel).
    The plot includes all analysis regions:
      the two common control regions (left)
      and the control, validation, and signal regions from the lep-had channel (middle) and from the had-had channel (right).
    The pulls in the control regions are small %
      by construction as the normalization factors obtained from the fit are applied.
    The hatched band gives the total statistical and systematic uncertainty in the background estimate in each region.
    The contribution of \ttbar events to CR \ttV and CR \VV is below a percent and not drawn here.
  }
  \label{fig:pullplot}
\end{figure}

The analysis results are summarized in \fig{fig:pullplot}, 
  which shows the data yields ($N_\text{obs}$) and background expectations ($N_\text{exp}$) in all analysis regions,
  and the resulting pulls $(N_\text{obs} - N_\text{exp}) / \sigma_\text{exp}$
  in the validation and signal regions,
  where $\sigma_\text{exp}$ includes the total uncertainty in the background estimate and the Poisson uncertainty in the data yield. %
The pulls in all but one validation region are below one standard deviation.
In the VR targeting \ttbar events with a real tau lepton in the lep-had channel, 
  an upwards fluctuation of around $2.3$ standard deviations is observed.
However, the distribution of \mttwo in this VR (top left plot in \fig{fig:CRVR_lephad})
  shows that the excess is confined to the single bin %
  farthest away from the signal region ($\GeV{60} < \mttwoLT < \GeV{80}$),
  and therefore inconsistent with a signal.

\subsection{Interpretation}

\begin{figure}[tbp]
  \centering
  \includegraphics[width=0.85\textwidth]{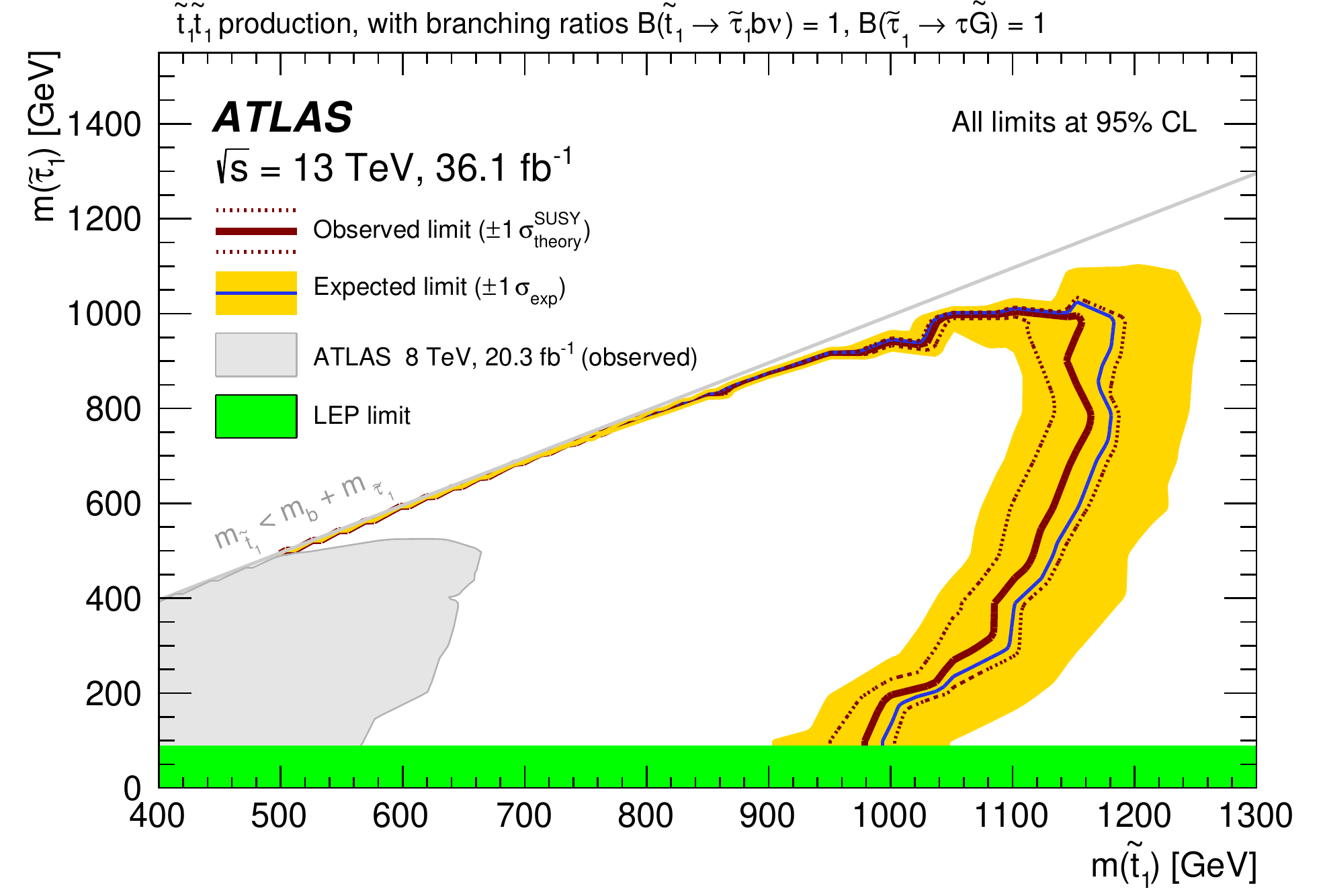}
  \caption{\discussID{7}
    Expected (solid blue line) and observed (solid red line) exclusion-limit contours at \percent{95} confidence level
      in the plane of top-squark and tau-slepton mass 
      for the simplified model, %
      obtained from the statistical combination of the lep-had and had-had channels,
      using full experimental and theory systematic uncertainties
      except the theoretical uncertainty in the signal cross section.
    The yellow band shows one-standard-deviation variations around the expected limit contour.
    The dotted red lines indicate how the observed limit moves 
      when varying the signal cross section up or down by the corresponding uncertainty in the theoretical value.
    For comparison, the plot also shows the observed exclusion contour from the ATLAS Run-1 analysis~\cite{SUSY-2014-04} as the area shaded in gray
      and the limit on the mass of the tau slepton (for a massless LSP) from the LEP experiments~\cite{web:LEP} as a green band.
  }
  \label{fig:Exclusion}
\end{figure}

In the absence of a significant excess beyond the SM prediction in either signal region,
  an exclusion limit is derived
  on the masses of the particles in the simplified signal model.
In contrast to the background fit, 
  the combined likelihood fit that is performed to derive the model-dependent exclusion limits
  allows for signal contamination in the CRs and includes the signal region. %
The \CLs prescription~\cite{Read:2002hq}
  is used to derive the probability that the signal-plus-background hypothesis is compatible with the observation
  and to set lower limits on the masses of the supersymmetric particles.

\Fig{fig:Exclusion} shows the expected and observed exclusion-limit contours at \percent{95} confidence level (CL) %
  obtained from the statistical combination of the lep-had and had-had channels
  with full experimental and theory systematic uncertainties.
Top-squark masses up to \limittone and tau-slepton masses up to \limittauone are excluded,
  which improves on the previous result from the ATLAS analysis of \ifb{20} of LHC data at \eighttev~\cite{SUSY-2014-04}
  by almost a factor of two in both mass parameters.
The had-had channel has better sensitivity than the lep-had channel over the whole mass plane,
  but the combination helps to improve the sensitivity, in particular for large tau-slepton masses.
For low tau-slepton masses,
  the sensitivity decreases and the limit on the top-squark mass is lower than at higher tau-slepton masses
  because the tau leptons from the tau-slepton decay become less energetic,
  which reduces the acceptance of the analysis selection.
When evaluating the distribution of the test statistic used for the hypothesis tests with simulated pseudoexperiments instead of the asymptotic formulae,
  the observed excluded range of top-squark masses is reduced by up to \GeV{40}.

\begin{table}
  \centering
  \setlength{\tabcolsep}{0.0pc}
  \begin{tabular*}{\textwidth}{@{\extracolsep{\fill}}lccccc}
    \noalign{\smallskip}\hline\noalign{\smallskip}
    Signal channel                        & \visxsec [fb]  &  $S_\text{obs}^{95}$  & $S_\text{exp}^{95}$ & \CLb & $p(s=0)$ ($Z$)  \\
    \noalign{\smallskip}\hline\noalign{\smallskip}
    SR LH   & $0.15$ &  $5.4$ & $ { 4.5 }^{ +2.6 }_{ -1.5 }$ & $0.65$ & $ 0.32$~$(0.47)$ \\%
    SR HH   & $0.13$ &  $4.7$ & $ { 4.6 }^{ +2.5 }_{ -1.5 }$ & $0.52$ & $ 0.48$~$(0.05)$ \\
    \noalign{\smallskip}\hline\noalign{\smallskip}
  \end{tabular*}
  \caption{\discussID{8}
    Left to right: observed \percent{95} confidence level (CL) upper limits on the visible cross section
    (\visxsec) and on the number of
    signal events ($S_\text{obs}^{95}$ ).  The third column
    ($S_\text{exp}^{95}$) shows the expected \percent{95} CL upper limit on the number of
    signal events, given the expected number (and $\pm 1\sigma$
    excursions on the expectation) of background events.
    The last two columns
    indicate the %
    \CLb value, 
    \ie the CL observed for the background-only hypothesis, %
    and the discovery $p$-value ($p(s = 0)$)
    and the corresponding significance ($Z$).
  }
  \label{tab:ULS-lh-hh}
\end{table}

In addition to the model-dependent limits above,
  the analysis results are also interpreted in terms of model-independent upper limits
  on the number of events from non-Standard-Model processes in the signal region, $S_\text{obs}^{95}$.
Dividing this number by the integrated luminosity of the dataset gives an upper limit on the visible signal cross section, 
  \visxsec,
  defined as the product of acceptance ($A$), reconstruction efficiency ($\epsilon$) and signal cross section ($\sigma$).
The model-independent limits are derived from a fit that is similar to the background fit,
  as it assumes no contamination by a potential signal in the CRs,
  but it includes the signal region with the extrapolated background contributions and a signal of variable strength.
The model-independent limits are shown in \tab{tab:ULS-lh-hh} separately for the two channels,
  again computed using the \CLs~prescription.
The lep-had channel yields a slightly lower expected limit on the number of signal events than the had-had channel 
  despite the larger expected SM background because the total uncertainty is smaller.
On the other hand, the mild excess of observed events is larger in the lep-had channel,
  so that the observed model-independent limit is lower for the had-had channel than for the lep-had channel,
  and the $p$-value for the background-only hypothesis in the lep-had channel is smaller.

\FloatBarrier

\section{Conclusion}
\label{sec:conclusion}

In this \article, 
  a search is presented
  for the direct pair production of supersymmetric top squarks in final states with two tau leptons, 
  jets identified as originating from $b$-hadron decays,
  and missing transverse momentum. %

The search uses a dataset with proton--proton collisions at a center-of-mass energy of \tttev,
  which was recorded with the ATLAS detector at the Large Hadron Collider in 2015 and 2016
  and has a total integrated luminosity of \intlumi.
Two exclusive channels are considered,
  which select events with 
    either two hadronically decaying tau leptons
    or one hadronically decaying tau lepton and one electron or muon.
Good agreement between the Standard Model prediction and the event yield observed in data is found in the signal region of each channel.
The analysis results are therefore interpreted in terms of upper limits on the production of supersymmetric particles.
In a simplified model with production of two top squarks, 
  each decaying via a tau slepton to a nearly massless gravitino as the lightest supersymmetric particle,
  masses up to $\massof{\tone}=\limittone$ and $\massof{\tauone}=\limittauone$ are excluded at \percent{95} confidence level,
  improving on previous limits in this model by almost a factor of two.
Model-independent limits allow the exclusion of visible cross sections of \unit[0.15 (0.13)]{fb} in the lep-had (had-had) channel
  for production of events beyond the Standard Model in this final state.

\ifisconf
\clearpage
\printbibliography
\else
\clearpage
\section*{Acknowledgments}
We thank CERN for the very successful operation of the LHC, as well as the
support staff from our institutions without whom ATLAS could not be
operated efficiently.

We acknowledge the support of ANPCyT, Argentina; YerPhI, Armenia; ARC, Australia; BMWFW and FWF, Austria; ANAS, Azerbaijan; SSTC, Belarus; CNPq and FAPESP, Brazil; NSERC, NRC and CFI, Canada; CERN; CONICYT, Chile; CAS, MOST and NSFC, China; COLCIENCIAS, Colombia; MSMT CR, MPO CR and VSC CR, Czech Republic; DNRF and DNSRC, Denmark; IN2P3-CNRS, CEA-DRF/IRFU, France; SRNSFG, Georgia; BMBF, HGF, and MPG, Germany; GSRT, Greece; RGC, Hong Kong SAR, China; ISF, I-CORE and Benoziyo Center, Israel; INFN, Italy; MEXT and JSPS, Japan; CNRST, Morocco; NWO, Netherlands; RCN, Norway; MNiSW and NCN, Poland; FCT, Portugal; MNE/IFA, Romania; MES of Russia and NRC KI, Russian Federation; JINR; MESTD, Serbia; MSSR, Slovakia; ARRS and MIZ\v{S}, Slovenia; DST/NRF, South Africa; MINECO, Spain; SRC and Wallenberg Foundation, Sweden; SERI, SNSF and Cantons of Bern and Geneva, Switzerland; MOST, Taiwan; TAEK, Turkey; STFC, United Kingdom; DOE and NSF, United States of America. In addition, individual groups and members have received support from BCKDF, the Canada Council, CANARIE, CRC, Compute Canada, FQRNT, and the Ontario Innovation Trust, Canada; EPLANET, ERC, ERDF, FP7, Horizon 2020 and Marie Sk{\l}odowska-Curie Actions, European Union; Investissements d'Avenir Labex and Idex, ANR, R{\'e}gion Auvergne and Fondation Partager le Savoir, France; DFG and AvH Foundation, Germany; Herakleitos, Thales and Aristeia programmes co-financed by EU-ESF and the Greek NSRF; BSF, GIF and Minerva, Israel; BRF, Norway; CERCA Programme Generalitat de Catalunya, Generalitat Valenciana, Spain; the Royal Society and Leverhulme Trust, United Kingdom.

The crucial computing support from all WLCG partners is acknowledged gratefully, in particular from CERN, the ATLAS Tier-1 facilities at TRIUMF (Canada), NDGF (Denmark, Norway, Sweden), CC-IN2P3 (France), KIT/GridKA (Germany), INFN-CNAF (Italy), NL-T1 (Netherlands), PIC (Spain), ASGC (Taiwan), RAL (UK) and BNL (USA), the Tier-2 facilities worldwide and large non-WLCG resource providers. Major contributors of computing resources are listed in Ref.~\cite{ATL-GEN-PUB-2016-002}.

 \clearpage
\printbibliography %
\clearpage

\begin{flushleft}
{\Large The ATLAS Collaboration}

\bigskip

M.~Aaboud$^\textrm{\scriptsize 34d}$,    
G.~Aad$^\textrm{\scriptsize 99}$,    
B.~Abbott$^\textrm{\scriptsize 124}$,    
O.~Abdinov$^\textrm{\scriptsize 13,*}$,    
B.~Abeloos$^\textrm{\scriptsize 128}$,    
S.H.~Abidi$^\textrm{\scriptsize 165}$,    
O.S.~AbouZeid$^\textrm{\scriptsize 143}$,    
N.L.~Abraham$^\textrm{\scriptsize 153}$,    
H.~Abramowicz$^\textrm{\scriptsize 159}$,    
H.~Abreu$^\textrm{\scriptsize 158}$
% ...
\bigskip
\\

% ...
$^{13}$Institute of Physics, Azerbaijan Academy of Sciences, Baku; Azerbaijan.\\
% ...
$^{34}$$^{(a)}$Facult\'e des Sciences Ain Chock, R\'eseau Universitaire de Physique des Hautes Energies - Universit\'e Hassan II, Casablanca;$^{(b)}$Centre National de l'Energie des Sciences Techniques Nucleaires (CNESTEN), Rabat;$^{(c)}$Facult\'e des Sciences Semlalia, Universit\'e Cadi Ayyad, LPHEA-Marrakech;$^{(d)}$Facult\'e des Sciences, Universit\'e Mohamed Premier and LPTPM, Oujda;$^{(e)}$Facult\'e des sciences, Universit\'e Mohammed V, Rabat; Morocco.\\
% ...
$^{99}$CPPM, Aix-Marseille Universit\'e, CNRS/IN2P3, Marseille; France.\\
% ...
$^{124}$Homer L. Dodge Department of Physics and Astronomy, University of Oklahoma, Norman OK; United States of America.\\
% ...
$^{128}$LAL, Universit\'e Paris-Sud, CNRS/IN2P3, Universit\'e Paris-Saclay, Orsay; France.\\
% ...
$^{143}$Santa Cruz Institute for Particle Physics, University of California Santa Cruz, Santa Cruz CA; United States of America.\\
% ...
$^{153}$Department of Physics and Astronomy, University of Sussex, Brighton; United Kingdom.\\
% ...
$^{158}$Department of Physics, Technion, Israel Institute of Technology, Haifa; Israel.\\
$^{159}$Raymond and Beverly Sackler School of Physics and Astronomy, Tel Aviv University, Tel Aviv; Israel.\\
% ...
$^{165}$Department of Physics, University of Toronto, Toronto ON; Canada.\\
% ...
$^{a}$ Also at  Department of Physics, University of Malaya, Kuala Lumpur; Malaysia.\\
$^{b}$ Also at Borough of Manhattan Community College, City University of New York, NY; United States of America.\\
$^{c}$ Also at California State University, East Bay; United States of America.\\
$^{d}$ Also at Centre for High Performance Computing, CSIR Campus, Rosebank, Cape Town; South Africa.\\
$^{e}$ Also at CERN, Geneva; Switzerland.\\
% ...
$^{*}$ Deceased

\end{flushleft}

 \fi

\end{document}